\documentclass[prd,onecolumn,showpacs,twoside,showpacs]{revtex4}

\usepackage{graphicx}
\usepackage{dcolumn}
\usepackage{bm}
\usepackage{color}
\usepackage{amsmath}
\usepackage{mathtools}

\begin{document}

\bibliographystyle{prsty}
\title{Hybrid method to resolve the neutrino mass hierarchy by
supernova (anti)neutrino induced reactions}
\author{D. Vale$^1$}
\email{dvale@phy.hr}
\author{T. Rauscher$^{2,3}$}
\author{N. Paar$^{3,1}$}
\email{npaar@phy.hr}
\affiliation{$^1$Department of Physics, Faculty of Science, University of Zagreb, 
Croatia}
\affiliation{$^2$Centre for Astrophysics Research, University of Hertfordshire, College Lane, Hatfield AL10 9AB, United Kingdom}
\affiliation{$^3$Department of Physics, University of Basel, Klingelbergstrasse 82, CH-4056 Basel, Switzerland}

\date{\today}
\begin{abstract}
We introduce a hybrid method to determine the neutrino mass hierarchy by simultaneous 
measurements of responses of at least two detectors to antineutrino 
and neutrino fluxes from accretion and cooling phases of core-collapse supernovae.
The (anti)neutrino-nucleus cross sections for $^{56}$Fe and $^{208}$Pb
are calculated in the framework of the relativistic nuclear energy density functional and weak interaction Hamiltonian, while the cross sections for inelastic scattering on free protons  
$\mathrm{p}(\bar{\nu}_\mathrm{e},\mathrm{e}^{+})\mathrm{n}$ are obtained 
using heavy-baryon chiral perturbation theory. 
The modelling of (anti)neutrino fluxes emitted from a protoneutron star in a core-collapse supernova 
include collective and {Mikheyev}-Smirnov-Wolfenstein effects inside the exploding star.
The particle emission rates from the elementary decay modes of the daughter nuclei are calculated for normal and
inverted neutrino mass hierarchy. It is shown that simultaneous use of (anti)neutrino detectors
with different target material 
allows to determine the neutrino mass hierarchy from the ratios 
of $\nu_\mathrm{e}$- and $\bar{\nu}_\mathrm{e}$-induced particle emissions. This hybrid method favors
neutrinos from the supernova cooling phase
and the implementation of detectors with heavier target nuclei ($^{208}$Pb) for the neutrino sector,
while for antineutrinos the use of free protons {in mineral oil or water}
is the appropriate choice.
\end{abstract}
\maketitle
\date{today}
\section{\label{sec:intro}Introduction}
%
%
Over the past years considerable
progress has been achieved in constraining the mixing parameters in the
neutrino oscillation framework \cite{Mak.62, Pon.67}, based on various experiments involving
atmospheric, solar, and terrestrial neutrinos \cite{Gon.08}. 
It is now well established that neutrinos have non-vanishing rest masses and that the flavor states
$\nu_e$, $\nu_{\mu}$, and $\nu_{\tau}$ are quantum mechanical mixtures of the 
vacuum mass eigenstates $\nu_1$, $\nu_2$, and $\nu_3$~ \cite{Stu.10}.
However, currently existing data do not determine the neutrino mass hierarchy, i.e.,
the sign of mass squared difference $\Delta m_{31}^2=m_3^2-m_{1}^2$. 
In the case of $\Delta m_{31}^2 >$ 0 one refers to the normal mass hierarchy
(NH), while $\Delta m_{31}^2 <$ 0 corresponds to the inverted mass 
hierarchy (IH). Although a number of techniques has been proposed to
resolve the neutrino mass hierarchy, to date this question still remains open 
and presents an important scientific challenge.
Recent approaches to resolve the neutrino mass 
hierarchy include methods based on reactor neutrinos \cite{Pet.02,Li.13,Cap.14},
various baseline experiments \cite{Ish.05}, Earth matter effects on the supernova
neutrino signal \cite{Lun.03,DDM.08}, the spectral swapping of supernova neutrino
flavors~\cite{Dua.07}, the rise time of supernova $\bar{\nu}_\mathrm{e}$ light 
curves \cite{Ser.12}, the analysis of meteoritic supernova material \cite{Mat.12},
and detection of atmospheric neutrinos in sea water or ice \cite{Win.13}.

The neutrino burst from a core collapsing supernova (ccSN) provides valuable information not only on the explosion mechanism,
but also about the neutrino mass hierarchy that is extremely difficult to determine in the laboratory~\cite{Raf.11}.
The first observation of the neutrino burst, from supernova SN1987A in the Large Magellanic Cloud, paved the way
for neutrino astronomy, and new frontiers in neutrino physics and astrophysics research.~\cite{Hir.87,Bio.87,Ale.88}.
Several supernova-neutrino detectors are currently running and a variety of new detectors are proposed to observe galactic ccSN burst~\cite{Sch.06,Sch.10}.
In the anticipation of the next galactic supernova (the occurrence rate is $\approx$1-3 events per century~\cite{Raf.11}), with expected
high statistics of (anti)neutrino induced detector events, in this paper we introduce a hybrid method
to resolve the neutrino mass hierarchy, based on neutrino and antineutrino 
reactions with  
$^{56}$Fe, $^{208}$Pb, and with free protons in terrestrial neutrino detectors.
The aim is to explore {novel perspectives for the implementation} of $\nu_\mathrm{e}$ and 
$\bar{\nu}_\mathrm{e}$ detectors, based on various
nuclei and/or protons (in mineral oil and water) as main target material.
Since for supernova SN1987A mainly the $\bar{\nu}_\mathrm{e}$ signal
has been detected, the role of {neutrino induced detector events for
understanding the fundamental properties of neutrinos remain largely unknown.}
While most of the supernova detectors {based on free protons or nuclear targets} 
still are primarily sensitive to antineutrinos, the recently developed
Helium and Lead Observatory (HALO) is sensitive
to neutrinos through charged current (CC) interaction mainly 
with $^{208}$Pb \cite{Dub.08}. 
{The hybrid method introduced in this work is based on the 
simultaneous implementation of two detectors, i.e., one for the neutrino and 
another one for the antineutrino signal from a supernova. Our model calculations
show that complementary information from two respective detectors can
provide constraints for the neutrino mass hierarchy.}

From supernova simulations \cite{Fis.10, Hud.11} and the neutrino signal analysis of 
SN1987A \cite{Lor.02, Vis.08} we understand that (anti)neutrino fluxes detected from a distance have two well distinguished components, i.e., one on a short time-scale ($\lesssim 1$ s) due to emission from the accreting 
matter onto the proton-neutron star and a
long-time scale component ($\approx 10$ s) 
due to Kelvin-Helmholtz cooling of 
the nascent proto-neutron star.
Thus, being able to trace 
neutrino and antineutrino fluxes from ccSN provide a
powerful tool to probe the dynamics of the stellar explosion but also various neutrino properties \cite{Sam.93, Das.10}. 
In particular, the flavor content of the neutrino signal evolves due to both neutrino collective effects and matter
 effects which can lead to a highly interesting interplay of these and distinctive spectral features \cite{Sig.93,Lun.13}. In dense neutrino gas collective oscillations occur due to nonlinear flavor evolution phenomena associated to neutrino-neutrino interactions~\cite{Peh.11,Peh.14}.
The time-dependent fluxes from ccSN can be used to improve the understanding 
of the neutrino flavor transformations in regions of high neutrino densities 
~\cite{Pan.92,Pan.92b,Sam.93,Dua.06,Dua.06b,Das.08,Das.10,Mir.11,Vaa.11}
and in transitions due to Mikheyev-Smirnov-Wolfenstein (MSW) 
effects occurring in the matter 
resonance layers of the stellar envelope \cite{Wol.79,Ful.87,Dig.00, Smi.03}.
More details on the neutrino flavor transformations in supernova are given
in recent reviews ~\cite{Dua.09,Dua.10}.
 
A microscopic description of neutrino-nucleus 
cross sections is crucial for the
interpretation of the experimental data. It is also important to guide the design of new 
neutrino detectors with sufficient energy and time resolution (and also different thresholds), 
which are sensitive enough to register charged current 
events in the energy region $E_{\nu_\mathrm{e}(\bar{\nu}_\mathrm{e})}\lesssim 30$ MeV from an intergalactic ccSN.
In this particular energy region the largest difference of 
(anti)neutrino spectra between two possible neutrino mass hierarchies is expected, due to
the impact of collective and MSW effects \cite{Sam.93,Das.10,Mir.11,Fog.12}.

Neutrino-nucleus interaction and cross sections in the energy region
of supernova neutrinos have already been studied within a
variety of microscopic approaches, including the shell
model~\cite{Hax.87,Eng.96,Hay.00,Sam.01,Suz.09}, the random phase 
approximation (RPA)~\cite{Aue.97,Sin.98,Vol.00,Vol.02}, 
continuum RPA (CRPA)~\cite{Kol.92,Kol.95,Jac.99,Jac.02,Bot.05},
combined CRPA and shell model~\cite{Aue.97,Kol.99,Kol.03,Vog.06},
the Fermi gas model~\cite{Wal.75,Gai.86,Kur.90}, quasiparticle 
RPA~\cite{Laz.07,Che.10,Cha.09,Tsa.11,Tsa.12}, projected QRPA~\cite{Sam.11}
and relativistic
quasiparticle RPA~\cite{Paa.08}.
In Ref.~\cite{Kol.01}
the cross sections for neutrino-induced particle emissions have been 
calculated for lead and iron neutrino detectors. Event rates from lead-based
supernova-neutrino detectors have also been studied in Ref.~\cite{Eng.03},
in order to explore the prospects for untangling the signatures of various
oscillation scenarios.

In the present study, a microscopic theory framework 
based on the relativistic nuclear energy density functional and weak 
Hamiltonian is employed in the description of charged-current neutrino-nucleus
reactions, including the properties of target nuclei,
neutrino induced excitations, and weak interaction transition 
matrix elements \cite{Paa.11,Paa.13}. 
In order to account for the $\nu_\mathrm{e}$($\bar{\nu}_\mathrm{e}$)- induced events in the detector, the primary
particle decay modes of the daughter nuclei are described, i.e., $\gamma$ decay and emission of one or two nucleons. 
Neutral current neutrino-nucleus reactions have not been taken into account in the present study. Their contributions in the neutron emission channel are small, less
than $\approx 10\%$ in heavy nuclei~\cite{Eng.03}.
To determine positron related signals in water and mineral oil,
$\bar{\nu}_\mathrm{e}$ cross sections of free protons were calculated in the framework introduced in
Ref.\ \cite{Rah.12, Rah.12e}, based on chiral perturbation theory.

One of the main advantages of the hybrid method established in this work
is the simultaneous detection of complementary parts of the neutrino spectra, i.e., $\nu_\mathrm{e}$ and $\bar{\nu}_\mathrm{e}$, in at least two different detectors, {thus providing} additional constraints as compared to traditional supernova analyses based {mostly on} antineutrino events in the detector.
The time separation and characteristics of the fluxes from different supernova phases 
are used for the prediction of the detector events. Although the absolute number of (anti)neutrino events seen by the detector depends on the distance between supernova and Earth, the presented hybrid method is almost independent on this parameter and offers a promising approach for solving the problem of the neutrino mass hierarchy.  

The paper is organized as follows. Section \ref{sec:SNneutrinos} includes the basic formalism 
and calculations of the (anti)neutrino fluxes incoming from an intergalactic ccSN, including the effects that depend on neutrino mass hierarchy. In Sec.\ \ref{sec:micro} we outline the nuclear models used for calculation of exclusive (anti)neutrino-nucleus cross sections and for the 
decay modes of daughter nuclei,  i.e., one- and two-particle emissions together with $\gamma$ decay. In Sec.\ \ref{sec:hybrid} we provide further details on the actual hybrid method to determine the neutrino mass hierarchy. 
The analysis of the detector responses to supernova (anti)neutrinos and results of the hybrid method are presented in Sec.\ \ref{sec:results}. This includes a discussion of the role of uncertainties in supernova (anti)neutrino fluxes and their impact on reactions in different types of detectors. Section \ref{sec:conclusion} summarizes the results and includes an outlook for future studies.

\section{\label{sec:SNneutrinos}Supernova neutrinos}

Neutrinos and antineutrinos are produced at various stages of the ccSN evolution~\cite{Jan.07}. Model simulations result in several 
characteristic phases of neutrino emission, and provide the evolution of the (anti)neutrino luminosities and of the average energies \cite{Jan.07,Hud.11}: (i) luminosity rise during core collapse, (ii) shock breakout burst, (iii) accretion phase in which matter is still falling onto the proto-neutron star (from few tens to few hundreds ms after bounce), ending when neutrino heating reverses the infall, and (iv) Kelvin-Helmholtz cooling of the hot proto-neutron star with a duration of 10 s or more, accompanied by mass outflow in the neutrino-driven wind. The initial neutrino bursts (i),(ii) are unlikely to be used for improved neutrino-signal analysis due to the small number of detector events expected for average intergalactic supernova distances \cite{Fis.10}. In the focus of the present study are neutrino induced responses from (anti)neutrinos associated to the accretion (iii) and cooling (iv) phases of ccSN.

During the accretion phase, the (anti)neutrino electron species with average energies $\langle E_{\nu_{e}}\rangle \approx 8-12$ MeV dominate~\cite{Cho.10, Fis.10, Hud.11}, with roughly a factor of two larger luminosity than non-electron neutrinos in \cite{Fis.10}, while in \cite{Hud.11} the difference is somewhat smaller. Within the cooling phase most of the energy released from the core leads to the creation of $\nu_\mathrm{e}\bar{\nu}_\mathrm{e}$ pairs, and neutrino luminosities may decrease by an order of magnitude.  As shown in supernova simulation, in the cooling phase the luminosities of electron and non-electron species decrease with time evolution, and their differences are small \cite{Fis.10, Hud.11}.
It is important to emphasize that the ordering of average energies remains the same through both phases, i.e., $\langle E_{\nu_e}\rangle\lesssim\langle E_{\bar{\nu}_e}\rangle\lesssim\langle E_{\nu_x}\rangle$. 
In this work, however, we will also explore 
deviations of luminosity ratios between different neutrino species 
to study the respective sensitivity of the
detector response in Sec.\ \ref{sec:results}.

In the inner region of the star, up to several tens of km above the neutrinospheres, 
neutrino flavor conversion is first frozen by synchronization due to the strength of 
neutrino-neutrino interactions \cite{Vaa.11,Mir.11}. At larger distances from the neutrinospheres, the interaction strength decreases and bipolar oscillations occur \cite{Sam.93, Dua.06, Das.10, Vaa.11}. Neutrino-neutrino interactions provide a nonlinear term in the equations
of motion \cite{Das.08, Mir.11} and the result of such collective effects
in the high-density neutrino region are spectral splits and swaps in the (anti)neutrino 
spectra. It is important to emphasize that recent studies of neutrino oscillations suggested that collective effects are strongly matter suppressed in the accretion phase \cite{Cha.11, Cha.14}.
In the present analysis we include the three known neutrino species ($\nu_\mathrm{e}$, ${\nu}_\mu$ and $\nu_\tau$), together with their antiparticles. We do not consider sterile types or a fourth generation. 

The matter density profile of the exploding star changes drastically along the neutrino path. At a specific density, neutrino interactions with matter start dominating through forward elastic neutrino-electron scattering and the Mikheyev-Smirnov-Wolfenstein
(MSW) effect induces efficient flavor transitions. There are two MSW resonance regions, the layer at higher densities (H-resonance layer) and lower densities (L-resonances layer)~\cite{Dig.00}. The neutrinospheres, however, are far below the resonance 
layers. In the neutrinosphere, the electron density is orders of magnitude larger 
than in the resonance layer. Thus, the flavor transition can simply be described by the Landau-Zener formula \cite{Dig.00, Smi.03}. In the iron core-collapse supernova the region dominated by collective effects and the two MSW resonant regions are spatially well 
separated (for neutrino energies above few MeV) \cite{Dig.00}. Therefore, probabilities 
for flavor transitions in different regions can be factorized~\cite{Vaa.11}. 
For the accretion case, the transition probability is given by a $3\times3$ matrix, defined as
\begin{equation}
P_\mathrm{a}= P_\mathrm{MSW}^\mathrm{L} P_\mathrm{MSW}^\mathrm{H}\;,
\end{equation}
due to matter suppression of collective effects, and for the cooling phase we get
\begin{equation}
P_\mathrm{c}= P_\mathrm{MSW}^\mathrm{L} P_\mathrm{MSW}^\mathrm{H} P_\mathrm{coll}\;,
\end{equation}
where $P_\mathrm{coll}$ represents the flavor transition probability matrix at the end of the collective region, while the matrices $P_\mathrm{MSW}^\mathrm{H}$ and $P_\mathrm{MSW}^\mathrm{L}$ are related to conversions in the high and low matter density resonant regions, respectively, due to the MSW effect. In general, matrix elements are different for neutrinos and antineutrinos, and hierarchy dependent. They were calculated in Refs. \cite{Dig.00, Smi.03, Vaa.11}. The arrival of the supernova shock in the outer layers of the star can leave a mark on the $\nu_\mathrm{e}$($\bar{\nu}_\mathrm{e}$) spectra and 
even cause a non-adiabatic conversion and multiple MSW effects \cite{Vaa.11}. In addition to a multi-zenith angle instability \cite{Ban.11}, it has recently been shown that a multi-azimuth angle instability can occur in the case of normal mass hierarchy \cite{Raf.13}. These effects were not 
taken into account in the supernova neutrino fluxes used here, i.e., a single-angle approximation was employed.

(Anti)neutrinos leave the surface of a star as mass eigenstates. 
On arrival at Earth, the incoming neutrino fluxes are incoherent due to spread of wave packets 
for typical supernova distances ($\approx 10$ kpc), thus suppressing oscillations along the travelled distance $L$.
In fact, they coincide with the 
neutrino fluxes on the surface of the star up to a geometrical factor $1/L^{2}$.
Few neutrinos, however, are 
detected as flavor states. The probability of detecting a particular flavor
specie is given by
\begin{equation}
P_\mathrm{a}^\mathrm{Earth} = P_\mathrm{PMNS} P_\mathrm{MSW}^\mathrm{L} P_\mathrm{MSW}^\mathrm{H}\;,
\end{equation}
for neutrinos emitted during the accretion phase, and 
\begin{equation}
P_\mathrm{c}^\mathrm{Earth} = P_\mathrm{PMNS} P_\mathrm{MSW}^\mathrm{L} P_\mathrm{MSW}^\mathrm{H} P_\mathrm{coll}\;,
\end{equation}
for emission in the cooling phase. The matrix $P_\mathrm{PMNS}$ represents absolute squares of the Pontecorvo-Maki-Nakagawa-Sakata matrix elements~\cite{Mak.62}. The above equations are applicable when the
detector is directly turned to the neutrino source, minimizing Earth effects on the (anti)neutrino spectra.
Therefore, the incoming fluxes are given by
\begin{equation}
F_\mathrm{a(c)}^\mathrm{Earth} = g(r) P_\mathrm{a(c)}^\mathrm{Earth} F_\mathrm{a(c)}^\mathrm{init}\;,
\end{equation}
where $g(r)=R_{\nu_{\alpha}}^2/r^2$ is a geometrical factor depending on the supernova-Earth distance $r$. This is a general expression and can even be used with time-dependent (anti)neutrino fluxes. The initial energy distribution for each flavor is given by
\begin{equation}
F_\mathrm{a(c);\nu_{\alpha}}^\mathrm{init}\left(E_{\nu_{\alpha}}, R_{\nu_{\alpha}}\right) = \Phi_{\nu_{\alpha}}^\mathrm{init}\left(R_{\nu_{\alpha}}\right) \phi_{\nu_{\alpha}}\left(E_{\nu_{\alpha}}\right) \;,
\end{equation}
where $\Phi_{\nu_{\alpha}}^\mathrm{init}(R_{\nu_{\alpha}}) = L_{\nu_{\alpha}}/(4\pi R_{\nu_{\alpha}}^2 \langle E_{\nu_{\alpha}}^0\rangle)$ is the initial number flux of a particular species, $L_{\nu_{\alpha}}$ denotes
the neutrino luminosity, and $R_{\nu_{\alpha}}$ corresponds to the radius of respective neutrinosphere. The initial neutrino spectra can be well described using a modified power-law distribution, also known as {\it alpha-fit}~\cite{Raf.03},
 \begin{equation}
\phi_{\nu_{\alpha}}(E_{\nu_{\alpha}}) = N \left( \frac{ E_{\nu_{\alpha}}}{ \langle E_{\nu_{\alpha}}^{0} \rangle}\right)^{\beta_{\nu}} exp \left[-\left(\beta_{\nu} + 1\right) \frac{ E_{\nu}}{ \langle E_{\nu_{\alpha}}^{0} \rangle} \right] \;,   
\label{flux1}	
\end{equation}
{with the normalization constant}
 \begin{equation}
 N = \frac{\left(\beta_{\nu} + 1\right)^{\beta_{\nu} + 1}}{\langle E_{\nu}^0\rangle \Gamma\left(\beta_{\nu} + 1\right)} \;,
 \end{equation}
where $\Gamma$ denotes the Euler Gamma function, and $\beta_{\nu}$ is the pinching parameter. In general, 
$\beta_{\nu}$ is different for each neutrino species and varies in time. This parameter, however, is only weakly constrained by the SN1987A data. Only an average value for each phase of $\bar{\nu}_\mathrm{e}$ spectra can be derived from a fit to the data. Alternatively, theoretical results from supernova simulations (including the general time- and phase-dependence behavior) can be used instead.
In this work, the {\it alpha-fit} distribution is employed, rather than the Fermi-Dirac distribution, due to its better description of the high energy tails of the spectra \cite{Kei.03}. 
As shown in recent analysis of high-resolution neutrino spectra from a spherically symmetric supernova model~\cite{Tam.12}, for the purpose of signal forecast in different detectors 
reasonably good accuracy can be achieved by considering the {\it alpha-fit} distribution.
High energy tails of neutrino distributions are important because of the strong increase of neutrino-nucleus cross sections with neutrino energy and the possible dependence of neutrino-induced two-neutron emission rates in heavy nuclei on high-energy neutrino spectra. Such neutron emissions will be analyzed and discussed in Secs.\ \ref{sec:micro} and \ref{sec:results}.

Finally, we also define the relative luminosity as 
 \begin{equation}
 l_{\nu_{\alpha}}^\mathrm{a(c)}(\Delta t) = \frac{L_{\nu_{\alpha}}^\mathrm{a(c)}(\Delta t)}{L_{tot}^\mathrm{a(c)}(\Delta t)} \;,
 \end{equation}
in order to avoid an explicit dependence on the total luminosity. 
In Sec. \ref{sec:results} we explore the sensitivity of the numbers of neutrino induced events in detectors and respective quantities introduced within the hybrid method, on the ratio of relative luminosities, $l_{\nu_\mathrm{e}}:l_{\bar{\nu}_\mathrm{e}}:l_{\nu_\mathrm{x}}$.

\section{\label{sec:micro}(Anti)neutrino-nucleus cross sections}

The charged-current (CC) neutrino- and antineutrino-nucleus reactions are considered in two steps, i.e., primary reactions
\\
\begin{subequations}
\begin{equation}
\nu_{\alpha} + {_{Z}X} _N  \rightarrow {_{Z+1}X}^*_{N-1} + l_{\alpha}^- ,
\label{prima}
\end{equation}
\begin{equation}
\bar{\nu}_{\alpha} + {_{Z}X} _N \rightarrow {_{Z-1}X}^*_{N+1} + l_{\alpha}^+ ,
\label{primb}
\end{equation}
\end{subequations}
and decay channels of the daughter nucleus in the case of Eq.\ (\ref{prima})
\begin{equation}
^A_{Z+1} X^*_{N-1} \rightarrow
\begin{dcases} 
^{A} _{Z+1} X_{N-1} + \gamma\\
^{A-1} _{Z+1} X_{N-2} + \mathrm{n}\\
^{A-1} _{Z} X_{N-1} + \mathrm{p}\\
^{A-4} _{Z-1} X_{N-3} + \alpha \\ 
^{A-2} _{Z+1} X_{N-3} + 2\mathrm{n}
\end{dcases}
\label{eq:seka}
\end{equation}
or in the case of Eq.\ (\ref{primb})
\begin{equation}
^A_{Z-1} X^*_{N+1} \rightarrow
\begin{dcases} 
^{A} _{Z-1} X_{N+1} + \gamma\\
^{A-1} _{Z-1} X_{N} + \mathrm{n}\\
^{A-1} _{Z-2} X_{N+1} + \mathrm{p}\\
^{A-4} _{Z-3} X_{N-1} + \alpha \\
^{A-2} _{Z-1} X_{N-1} + 2\mathrm{n}
\end{dcases}
\label{eq:sekb}
\end{equation}
where $l_{\alpha}$ denotes the charged lepton (electron, muon, or its antiparticle). In this paper, however, we are mainly focused on electron species. The detector events following charged-current (CC) primary reactions with muon (anti)neutrinos are negligible for typical ccSN fluxes. 
The CC (anti)neutrino-nucleus cross section is given by \cite{Con.72,Wal.75}
\begin{equation}
\label{CS}
\left (  \frac{d \sigma_{ \nu_{\alpha} (\bar{\nu}_{\alpha})  }   }{ d \Omega }   \right ) =
 \frac{1 }{ (2 \pi )^2}V^2 p_l E_l
 \sum_{\begin{minipage}{0.7cm} \scriptsize lepton \\  spins \end{minipage} } \frac{ 1}{ 2J_i + 1}   \sum_{M_i M_f} | \langle
f |  \hat{H}_{W} |i  \rangle | ^2 \;,
\end{equation}
where $p_l$ and $E_l$ are the momentum and energy of the outgoing lepton,
respectively, and $V$ is a normalization volume. The Hamiltonian $ \hat{H}_{W} $ of the weak interaction 
is expressed in the standard current-current form, i.e., in terms of the 
nucleon $\mathcal{J}_{ \lambda }(\bm{x})$ and lepton $j_{ \lambda }(\bm{x})$ 
currents, as
\begin{equation}
\hat{H}_{W}=- \frac{G }{ \sqrt{2} }  \int d\bm{x}
\mathcal{J}_{ \lambda }(\bm{x})j^{ \lambda }(\bm{x}) \;,
\end{equation}
and the transition matrix elements are
\begin{equation}
\langle f | \hat{H}_{W}| i \rangle = -  \frac{ G}{\sqrt{2}}l_{ \lambda }
 \int d \bm{x} e^{- i \bm{q} \bm{x}}
\langle f | \mathcal{J}^{ \lambda }(\bm{x}) | i \rangle \; .
\end{equation}
For the purpose of this work, the cross sections for the 
charged current $\nu_{e}(\bar{\nu}_{e})$-nucleus reactions are calculated for the following target 
nuclei: $^{12}$C, $^{16}$O, $^{56}$Fe, and $^{208}$Pb. 
The exclusive cross sections of the primary reactions are described in the
framework based on the relativistic nuclear energy density functional (RNEDF) \cite{Paa.08,Paa.11,Paa.13}, 
by employing the density dependent effective interaction DD-ME2 \cite{Lal.05} 
in the particle-hole channel, while pairing correlations are described by the
pairing part of the finite-range Gogny interaction with set D1S \cite{BGG.91}.
This framework has been successful in the description of Gamow-Teller transitions
and other charge-exchange modes, thus it provides a reasonable method
of choice to calculate relevant neutrino-induced transitions in nuclei \cite{Paa.11, Niu.11}.
Transition matrix elements for neutrino-induced reactions are obtained using 
the general formalism from Refs.~\cite{Con.72, Wal.75}. 
This method allows to determine the ground state properties of 
target nuclei and transitions induced by neutrinos in a consistent way.

The exclusive cross sections are calculated as functions of excitation energy of the initial nuclei, 
including all contributions from the initial ground state of the even-even nucleus to the 
particular excited state of the daughter odd-odd nucleus, for all relevant multipolarities $J\leq 5$ and both parities. With the incoming (anti)neutrino fluxes in a terrestrial detector, the transition cross section to a particular state $i$ of the daughter nucleus is given by
\begin{equation}
\sigma_{i} = \int\limits_{E_{\nu_{\alpha}} \geq E_\mathrm{thresh}}
\sigma_{i}\left(E_{\nu_{\alpha}}\right)
F^\mathrm{Earth}_{\nu_{\alpha}} \left(E_{\nu_{\alpha}}\right) dE_{\nu_{\alpha}} \;,
\end{equation}
and the flux averaged cross section is
\begin{equation}
\langle\sigma_{i}\rangle = \frac{\sigma_{i}}{F^\mathrm{tot}_{\nu_{\alpha}}} \;,
\end{equation}
where the index $i$ denotes the excited state in the nuclear daughter with excitation energy 
$E^\mathrm{excit}_i$, spin $J$, and parity $\pi$. The quantity $E_\mathrm{thresh}$ is the energy threshold for a particular reaction (they usually vary from a few hundred keV to $\approx 10$ MeV). 

The primary reactions create daughter nuclei which de-excite through emission of photons or particles. Only these emissions can be registered in a terrestrial detector.
Since we want to explore signatures of neutrino-induced reactions in such detectors, the emission channels of the daughter nuclei, as shown in Eqs.\ (\ref{eq:seka}) and (\ref{eq:sekb}), have to be followed in a calculation. In this calculation we included all the channels shown in the above equations but here we particularly focus on neutron emission, as charged particle emission is negligible in comparison. The neutrino flux-integrated emission cross section $\sigma_{i\kappa}=\sigma_i P_{i\kappa}$ for daughter state $i$ and a particular channel $\kappa$ is simply the product of the flux-integrated cross section $\sigma_i$ and a branching ratio $P_{i\kappa}$, describing the relative contribution of channel $\kappa$ within all possible decay channels. This branching was calculated from the ratio of transmission coefficients $T_{i\kappa}$ obtained from solutions of the time-independent Schr\"odinger equation in the optical model,
\begin{equation}
\label{eq:trans}
P_{i\kappa}=\frac{T_{i\kappa}}{\sum_{\kappa'} T_{i\kappa'}}\;, \kappa,\kappa' \neq 2\mathrm{n} \;.
\end{equation}
These transmission coefficients are related to reaction widths of resonances and are similar to those used in the Hauser-Feshbach model which describes compound nucleus reactions \cite{HauF52}. Therefore, we used the Hauser-Feshbach code SMARAGD \cite{SMARAGD,Raus11} to determine $P_{i\kappa}$. The procedure to calculate transmission coefficients is similar to the one described in \cite{Raus11,Raus00}. Emissions to final states include emissions to experimentally known discrete levels up to a final state excitation energy $E_\mathrm{fin}$. Above $E_\mathrm{fin}$ an integration over a nuclear level density for spins $J\leq 12$ and both parities is invoked. Thus, the calculated emission spectrum contains both discrete and continuous contributions, as is also typical for compound nucleus reactions. Here, the ``compound'' nucleus is formed by the primary CC weak reactions. The same default settings of the code were used as in \cite{Raus00} but with updated experimental levels (up to 40 known excited states from the 2010 version of NuDAT \cite{nudat}) and level density (including parity dependence as described in \cite{darko07}). Emission thresholds are given by the particle separation energies $S_\kappa$ in the emitting nucleus. These were computed from measured nuclear mass differences taken from \cite{AME}. It should be noted that the $\gamma$ cascade in the daughter was not followed in detail and therefore neutron emission after $\gamma$ deexcitation(s) is not included. We expect it to be negligible, anyway, because significant contributions to the total particle emissions only come from $\sigma_i$ with moderate excitation energy, $E_i^\mathrm{excit}\lesssim 25$ MeV, and separation energies are of the order of 10 MeV.

For emission of the second neutron, $T_{i\,\mathrm{n2}}=T_{i\rightarrow j\,\mathrm{n}}\sum_{j_{i\rightarrow j}}P_{j\,\mathrm{n}}^*$ was calculated from a recursive procedure, starting from the transmission coefficient for one-neutron emission from state $i$ to final state $j$, $T_{i\,\mathrm{n}}$. For all final states $j$ possibly populated by emission of one neutron, the procedure of determining a branching ratio $P^*$ for single-neutron emission from the new daughter nucleus in state $j$ was repeated.

Using the above approach, it is not only possible to calculate flux-integrated cross sections for the individual particle emission channels but it is also possible to determine the emission spectrum $\mathcal{S}_\kappa (E_\mathrm{em})$, i.e., the number of particles emitted with a given energy $E_\mathrm{em}$. This is useful if the detector cannot only provide an energy-integrated signal but also resolve the actual particle spectrum. To illustrate a first application of the hybrid method, however, in the following we restrict ourselves to predicting particle counts in the detector, i.e., energy integrated results. The detailed emission spectra are available from the authors on request and may provide further constraints in future analyses of neutrino signatures in detectors. Since the studies of lead detector efficiency are still in 
progress~\cite{Sch.15}, in the present analysis perfect detector response is assumed, with the efficiency denoted by $\eta \to 1$.

The rate of particles registered in the detector is given by
\begin{equation}
\frac{dN_\kappa}{dt} = N_\mathrm{T} D_{\kappa} \sum_i \sigma_i \widetilde{P}_{i \kappa} \;,
\label{kappaN}
\end{equation}
where in the case of emission of only one neutron (1n) from daughter nucleus we 
have $\widetilde{P}_{i1n} = P_{i1n}-P_{i2n}$, while for $\gamma$, one-proton (1p), two-neutron (2n) and  emission of $\alpha$ particle $\widetilde{P}_{i \kappa}={P}_{i \kappa}$.
The total number of targets (nuclei or protons) in the detector is $N_\mathrm{T} = X m_\mathrm{d}N_\mathrm{A}/M_r$,
{where $m_\mathrm{d}$ denotes the detector mass, $X$ the number of particular targets (e.g., $X=2$ corresponds to the number of protons in water molecule, otherwise if no molecules are formed, $X=1$), $M_r$
the respective molar weight, and $N_\mathrm{A}$ the Avogadro constant (see Sec.\ \ref{sec:hybrid} for more details).

In addition to $\nu_\mathrm{e}(\bar{\nu}_\mathrm{e})$-nucleus cross sections, in this work also inelastic $\bar{\nu}_\mathrm{e}$ scattering on free protons is taken into account. This is relevant for detectors using mineral oil (alkanes, which contain CH$_{2}$ groups)
or water (H$_2$O). In the calculation, we employ heavy-baryon
chiral perturbation theory, which also
includes radiative corrections ~\cite{Rah.12, Rah.12e}. 
The calculated cross sections for inelastic $\bar{\nu}_\mathrm{e}+\mathrm{p}$ 
scattering are shown in Fig.~\ref{anu_proton}. In Sec.\ref{sec:results}, this cross section is employed in 
modeling the number of the detector events induced by antineutrinos in mineral oil and water.

\section{\label{sec:hybrid}Description of the hybrid method}

Predictions of the absolute values of the number of events for NH or IH may not provide 
conclusive information on the neutrino mass hierarchy due to the overlaps of the expected value ranges caused by the considerable uncertainties in the modeling of supernova (anti)neutrino fluxes~\cite{Hud.11}.
In the following we introduce a hybrid method that could constrain the neutrino mass
hierarchy. It is based on observing CC reactions with at least two detectors which cover complementary parts of supernova (anti)neutrino spectra, i.e., one detector is sensitive mainly to
$\nu_\mathrm{e}$ while the other to $\bar{\nu}_\mathrm{e}$.
At the core of the hybrid method is the definition of 
a set of hierarchy-dependent quantities, involving ratios of event numbers from several detectors and making use of different reactions as well as different phases of supernova evolution.
In this way the dependence on the absolute values of some of the involved quantities is reduced, e.g., on the luminosities, on the supernova-Earth distance, and on the mass of the detector. In this work, we explore the feasibility to provide conclusive information on neutrino mass hierarchy by using
the relative quantities of such a hybrid method based on two detectors, rather than depending on the absolute number of the events in a single detector.

In the analysis of time-dependent detector events, for each reaction channel the ratio of the numbers of emitted particles, both for the supernova accretion and cooling phase, is given by,
\begin{equation}
\label{c}
R_{\kappa_\mathrm{A} \kappa_\mathrm{B}}^\mathrm{a(c);AB}(\Delta t_{j}) = \frac{\sum_{\tau_\mathrm{A}} N_{\kappa_\mathrm{A},\bar{\nu}_\mathrm{e}}^{\mathrm{a(c)};\tau_\mathrm{A}}(\Delta t_{j})}{\sum_{\tau_\mathrm{B}} N_{\kappa_\mathrm{B}, \nu_\mathrm{e}}^{\mathrm{a(c)};\tau_\mathrm{B}}(\Delta t_{j})} \;,
\end{equation}
and similarly its time-integrated variant. 
The indices A and B identify the detector with its particular decay or deexcitation channel $\kappa_\mathrm{A}$, $\kappa_\mathrm{B}$, of the daughter nucleus (see Sec.\ \ref{sec:micro}). In addition, the labels $\tau_A$ and $\tau_B$ identify whether nuclei or free protons are involved in CC reactions in the detectors. For example, there are several different target nuclei in a water or mineral oil detector (free protons and isotopes of C or O), which have different contributions to the number of CC events due to their different CC reaction
cross sections. We assume that the signal of a particular emission mode can be well distinguished from signals of other modes. Furthermore, the time-bin $\Delta t_j$ should be wide enough to allow for good statistics of the events and it should be similar (comparable) for both detectors.
If the $\nu$-$\nu$ interaction is sufficiently strong to cause collective effects in the region up to few hundred km above neutrinosphere, then flavor transitions will occur in complementary parts of the spectra, causing spectral splits and swaps. Furthermore, in the case of NH the MSW effect in both resonant regions will contribute to transitions between neutrinos, while the case of IH is characterized by transitions in the lower resonant region only. This is different, however, for antineutrinos which experience transitions in the low resonant region for NH and in both resonant regions for IH (for details, see \cite{Dig.00}). The quantity $R_{\kappa \lambda}^{AB}$ is a relative measure of total supernova effects on initial (anti)neutrino spectra.

From Eqs.\ \eqref{kappaN} and \eqref{c} we get
\begin{equation}
R_{\kappa_\mathrm{A} \kappa_\mathrm{B}}^\mathrm{a(c);AB}(\Delta t_{j}) = \Lambda_\mathrm{AB} \frac{D_{\kappa_\mathrm{A}} \sum\limits_{i, \tau_\mathrm{A}} X_\mathrm{A}^{\tau_\mathrm{A}}\sigma_{i(\bar{\nu}_\mathrm{e})}^{\tau_\mathrm{A}}(\Delta t_j) \widetilde{P}_{i \kappa_\mathrm{A}}^{\tau_\mathrm{A}} }{D_{\kappa_\mathrm{B}} \sum\limits_{i, \tau_\mathrm{B}} X_\mathrm{B}^{\tau_\mathrm{B}} \sigma_{i(\nu_\mathrm{e})}^{\tau_\mathrm{B}}(\Delta t_j) \widetilde{P}_{i \kappa_\mathrm{B}}^{\tau_\mathrm{B}} } \;,
\label{multi}
\end{equation}
where
\begin{equation}
\Lambda_\mathrm{AB} = \frac{\frac{m_\mathrm{A} N_\mathrm{A}}{M_r^\mathrm{A}}}{\frac{m_\mathrm{B} N_\mathrm{A}}{M_r^\mathrm{B}}} \;,
\end{equation}  
is constant for a given combination of neutrino (B) and antineutrino (A) detector. The coefficient $X^\tau_\mathrm{A(B)}$ stands for the number of targets (atoms or protons) in a molecule for mineral oil or water detectors. In the case of a detector using a metal, $X^\tau_\mathrm{A(B)} = 1$. 
The relative molecular (A) or atomic (B) molar weight is denoted by $M_r^\mathrm{A(B)}$. Note that even in a single detector several targets can give contributions to the total $\gamma$, 1n, 1p and 2n emissions. In Eq.\ \eqref{multi} one can see that $R_{\kappa_\mathrm{A}\kappa_\mathrm{B}}^\mathrm{AB}$ contains three types of information, the coefficient $\Lambda_\mathrm{AB}$ depends only on the combination of detectors used, the coefficients $D_\kappa$ and $\widetilde{P}_{i\kappa}^\tau$ are connected to the particle emission and its probability of observation, and third quantity $\sigma_i^\tau$, which is the transition probability at a given neutrino flux per target nucleus.
The latter term is rather complex as it depends on the structure of the nuclear ground and excited states, and on the (anti)neutrino spectra. Thus, it also contains imprints of collective and MSW effects, which are generally different for NH and IH.

In order to eliminate the dependence on the particular detector combination, thus using only information on particular targets involved in CC reactions for $\eta \rightarrow 1$ (perfect detector response), we define
the reduced quantity,
\begin{equation}
r_{\kappa_\mathrm{A} \kappa_\mathrm{B}}^\mathrm{AB} = \lim\limits_{\eta \rightarrow 1}R_{\kappa_\mathrm{A} \kappa_\mathrm{B}}^\mathrm{AB}/\Lambda_\mathrm{AB} \;,
\label{reducedc}
\end{equation}
which is an effective ratio of transformation probabilities (including decay modes) per single atom (B) or molecule (A) caused by complementary parts of the incoming supernova neutrino fluxes.  
The quantity $r_{\kappa_\mathrm{A} \kappa_\mathrm{B}}^\mathrm{AB}$ still contains information about targets included in the processes, and on the empirical structure for molecules when dealing with water and mineral oil. In other words, the reduced quantity $r_{\kappa_\mathrm{A} \kappa_\mathrm{B}}^\mathrm{AB}$ is good measure for hierarchy-dependent effects, i.e., a relative measure of the effects that shape (anti)neutrino spectra, independent of the absolute values of detector mass, effective number of targets, and explicit supernova-Earth distance. 

\section{\label{sec:results}Results}

The results of the present analysis are twofold: the first category is related to the detector response to the fluxes of supernova (anti)neutrinos, i.e., the induced particle emissions from target nuclei; the second is related to applications of the hybrid method to constrain the neutrino mass hierarchy.
\subsection{\label{sec:detect}Supernova (anti)neutrino fluxes}

In order to investigate the feasibility of the hybrid method despite of the current uncertainties in predicting neutrino fluxes~\cite{Hud.11},  the effects of various sets of initial supernova flux parameters are explored.
We used several combinations of initial luminosities in the accretion ($\sim 10^{52}$ erg/s) and 
cooling phase ($\sim 10^{51}$ erg/s) fluxes, with a fixed total emitted energy for each phase (see Tab.~\ref{table1}), i.e., $L_\mathrm{tot}^\mathrm{acc} \approx 0.4 \times 10^{53}$ erg for a duration of $0.4$ s, and $L_\mathrm{tot}^\mathrm{cool} \approx 2 \times 10^{53}$ erg for a duration of $10$ s. We also varied the initial average energies of neutrinos
from 8 to 12 MeV for $\nu_\mathrm{e}$, from 11 to 15 MeV for $\bar{\nu}_\mathrm{e}$, and
from 11 to 19 MeV for non-electron 
species. Thus, the range of values from different simulations of supernova neutrino fluxes is covered
\cite{Fis.10, Mir.11, Hud.11}, and the sensitivity of the hybrid method on the differences in 
neutrino fluxes can be explored.
In the present analysis, the initial averaged energies of neutrino species were kept in the canonical order, as shown in Sec.\ \ref{sec:SNneutrinos}, with two additional conditions: $\langle E_{\bar{\nu}_\mathrm{e}}^0 \rangle - \langle E_{\nu_\mathrm{e}}^0 \rangle \lesssim 3$ MeV and  $\langle E_{{\nu}_x}^0 \rangle - \langle E_{\bar{\nu}_\mathrm{e}}^0 \rangle \lesssim 4$ MeV. 
The pinching parameter $\beta_{\nu}$ in Eq.(\ref{flux1})
was set to 4.0 for the accretion and 3.0 for the cooling phase fluxes \cite{Mir.11,Raf.03}. 
In calculating transition probabilities beyond the collective region, we have used fixed values of splitting energies from a supernova neutrino flux analysis \cite{Cho.10}.
The incoming (anti)neutrino fluxes were calculated including 
collective and MSW effects in the core-collapsing star (see Sec. II).
Model calculations include best-fit values of neutrino oscillation parameters. They can be found, e.g., in \cite{Fog.12}, although somewhat different values are given in \cite{For.12, Gon.12}.
As a test case for ccSN in our galaxy, an imaginary star 25000 l.y. away from Earth was 
assumed. 

As an example, Fig.~\ref{fluxes} shows the set of incoming $\nu_e$ and $\bar{\nu}_e$ 
fluxes of the accretion and cooling supernova phases for the normal (NH) and inverted mass hierarchy (IH).
Three combinations of the luminosity ratios of (anti)neutrino species for type II supernova spectra
in accretion(A1-A3) and cooling (C1-C3) phase are considered, as given in Tab.~\ref{table1}.
The initial average energies of the supernova fluxes shown in Fig.~\ref{fluxes} are 
$\langle E^0_{\nu_\mathrm{e}} \rangle$ = 10 MeV, $\langle E^0_{\bar{\nu}_\mathrm{e}} \rangle$ = 13 MeV, and
$\langle E^0_{\nu_x} \rangle$ = 15 MeV.


\subsection{\label{sec:nuclresponse}Nuclear responses to supernova (anti)neutrinos}

Figures \ref{kombo_fe} and \ref{kombo_afe} show examples of the CC primary cross sections of $^{56}$Fe averaged over supernova cooling phase fluxes of $\nu_\mathrm{e}$ and $\bar{\nu}_\mathrm{e}$ (for the case C3 in Tab.\ \ref{table1}), respectively, and displayed as a function of excitation energy in the initial nucleus. 
Contributions to the cross sections are shown separately for several multipoles, from $J^{\pi}=0^{\pm}$ to $J^{\pi}=3^{\pm}$.
The energy threshold for CC reactions in $^{56}$Fe is relatively low for $\nu_\mathrm{e}$ (4.57 MeV) 
and $\bar{\nu}_\mathrm{e}$ (4.72 MeV).
The $\nu_\mathrm{e}$-$^{56}$Fe cross section is mainly determined by
$J^{\pi}=0^{+},1^{+}$ transitions, while the contribution of $J^{\pi}=1^{-},2^{-}$ states
is an order of magnitude smaller. Other multipolarities contribute only marginally. 
As shown in the upper panel of Fig.~\ref{kombo_afe}, the $\bar{\nu}_\mathrm{e}$-$^{56}$Fe reaction
cross section, on the other hand, is dominated by $J^{\pi}=1^{+}$ transitions, with a major contribution from
the excited state at 4.5 MeV. The lower panels of Figs.\ \ref{kombo_fe} and \ref{kombo_afe} show the ratio of flux averaged cross sections between two hierarchies for the specific incoming cooling phase $\nu_\mathrm{e}$ or $\bar{\nu}_\mathrm{e}$ flux C3. They indicate that exclusive transitions to particular states should be reduced
in the case of IH by factors $\approx 1.1-2.15$ for $\nu_\mathrm{e}$ and $\approx 1.15-1.32$ for $\bar{\nu}_\mathrm{e}$. A similar effect is expected for the other flux cases. 

The total $\nu_\mathrm{e}$-nucleus CC reaction cross section is increasing rapidly with 
increasing number of neutrons in the nucleus~\cite{Paa.13}. 
The relative difference in the depths of the proton 
and the neutron part of the mean field potential increases in nuclei with neutron excess, however, Pauli blocking strongly suppresses transitions from proton to neutron 
quasiparticle states in $\bar{\nu}_\mathrm{e}$-nucleus reactions. 
This effect is evident in the $\bar{\nu}_\mathrm{e}$-$^{56}$Fe reaction, where 
the cross sections are an order of magnitude smaller
than in the case of neutrino induced reaction (see Figs.~\ref{kombo_fe} and \ref{kombo_afe}), 
although $^{56}$Fe has a relatively small neutron excess $(N-Z=4)$.
The upper panel of Fig.~\ref{kombo_pb} shows the exclusive cross sections for the $\nu_\mathrm{e}$-$^{208}$Pb 
reaction, illustrating an example of a very neutron-rich target. The 
overall cross sections are considerably larger than for $^{56}$Fe, thus
indicating that a lead-based detector is a reasonable choice to efficiently
detect supernova neutrinos. On the other hand, due to the strong blocking effect in
the single-particle spectra, $\bar{\nu}_\mathrm{e}$-$^{208}$Pb cross sections are considerably 
reduced, i.e., the lead detector allows measurements of supernova-neutrino induced 
events only.

\subsection{\label{sec:detectresponse}Detector responses to supernova (anti)neutrinos}

By employing the framework outlined in Sec.~\ref{sec:micro}, we study neutrino induced events in detectors based on various target material. Since perfect efficiency is assumed, we note that in the realistic case the observed number of the events may considerably reduce, e.g., the efficiency to detect a neutron generated in a lead-based detector can be as low as a few 
tens of percent~\cite{Sch.12,Sch.15}. Furthermore, it is assumed that the detector
is turned directly to the incoming neutrino flux, thus 
eliminating Earth effects on neutrino spectra. 
Four cases of target material were considered, mineral oil ($\text{CH}_{2}$), water (H$_{2}$O), $^{56}$Fe, and $^{208}$Pb. The mass of the detector was taken to be 1 kt for all cases of target material.

The numbers of (anti)neutrino-induced events due to 1n, 2n, and $e^+$ emissions
for several different accretion supernova fluxes are shown in Tab.~\ref{table2}, both for 
normal and inverted neutrino mass hierarchies. Three combinations of the luminosity ratios (A1,A2,A3) are considered, as given in Tab.~\ref{table1}. In addition, three sets of the
initial average energies of the supernova fluxes are used, 
 (i) $\langle E_{\nu_\mathrm{e}}^0 \rangle$ = 8 MeV, $\langle E_{\bar{\nu}_\mathrm{e}}^0 \rangle$ = 11 MeV, $\langle E_{\nu_x}^0 \rangle$ = 13 MeV; (ii) $\langle E_{\nu_\mathrm{e}}^0 \rangle$ = 10 MeV, $\langle E_{\bar{\nu}_\mathrm{e}}^0 \rangle$ = 13 MeV, $\langle E_{\nu_x}^0 \rangle$ = 15 MeV; and (iii) $\langle E_{\nu_\mathrm{e}}^0 \rangle$ = 12 MeV, $\langle E_{\bar{\nu}_\mathrm{e}}^0 \rangle$ = 15 MeV, $\langle E_{\nu_x}^0 \rangle$ = 19 MeV.
The number of $\bar{\nu}_\mathrm{e}$-related events, i.e., the number of emitted positrons, is expected to depend not only on the initial average energies of the $\nu$ spectra, but also on the initial luminosity ratios between neutrino species. Due to large overlap of the respective positron events, however, it is difficult to put any constraint on the neutrino mass hierarchy from these events alone.

The number of single-neutron emissions in $^{56}$Fe ($^{208}$Pb) is $\approx 10$ (few) times smaller than positron events related with free protons in water and mineral oil, while two-neutron events in $^{208}$Pb are negligible for cases (i) and (ii). The number of single-neutron events in $^{56}$Fe and $^{208}$Pb significantly increases with the initial average energies of the $\nu$ spectra. The number of emitted neutrons, however, remains ambiguous for the shown cases, due to the strong matter suppression of collective effects, which was also observed in \cite{Vaa.11}. Therefore, the analysis of the detector response to the (anti)neutrinos from the accretion phase does not allow to distinguish between the two neutrino mass hierarchies.

Table~\ref{table3} shows the number of detector events in the same target materials as discussed above, but for the (anti)neutrino fluxes from the cooling phase, for the luminosity ratios in Tab.~\ref{table1}, denoted as C1,C2,C3. Again, three sets of the initial average energies of the supernova fluxes are considered.
The number of positron events in water or mineral oil induced by $\bar{\nu}_\mathrm{e}$ is at least a few times larger than the number of primary $\nu_\mathrm{e}$ reactions with $^{56}$Fe and $^{208}$Pb for all investigated cooling phase neutrino fluxes (with 
different values of initial average energies as shown in the table). 
The number of single-neutron emissions for $^{56}$Fe and for one- and two-neutron emissions for $^{208}$Pb target
is larger in the case of NH (see Figs.\ \ref{kombo_fe} and \ref{kombo_pb}). 
Contrary to the accretion phase, the neutrino fluxes from the cooling phase result in an increased number of the events for NH and thus in a reasonable separation between the numbers of the particle emissions from nuclei for the two hierarchies.

The most pronounced response to CC interactions with $\nu_\mathrm{e}$ 
is obtained for $^{208}$Pb. Due to large number of protons $(Z=82)$, 
Coulomb effects enhance the phase 
space for emitted electrons \cite{Ful.99}. It also has large neutron excess $(N-Z=44)$ 
and relatively low threshold (2.88 MeV) for CC reactions.
For the supernova cooling phase, we obtained a flux averaged cross section 
for neutron emission from $^{208}$Bi of $\approx 10^{-40}$ cm$^2$. 
In addition, a significant contribution of two-neutron events was obtained due to relatively high two-neutron cross sections ($\approx 10^{-41}-10^{-40}$ cm$^2$) at energies above the two-neutron separation energy of 23 MeV relative to the mother nucleus (see upper panel of Fig.\ \ref{kombo_pb}).
The same applies for all sets (i)-(iii) of neutrino fluxes.
The sensitivity of the number of 2n-emissions on the type of mass hierarchy is evident 
in the cooling phase, i.e., the ratio $N_\mathrm{2n}^\mathrm{NH}/N_\mathrm{2n}^\mathrm{IH}\approx2$ (see lower panel of Fig.~\ref{kombo_pb} and Tab.~\ref{table3}).

The cross sections for $\bar{\nu}_\mathrm{e}-^{208}$Pb reactions are strongly 
suppressed (by three orders of magnitude) due to Pauli blocking of the neutron
single-particle states. 
Thus, lead presents an excellent choice for a supernova neutrino detector to be used in 
the hybrid method. In order to obtain good statistics of the events, though, a sufficient amount of target material has to be available, e.g., $\approx 1$ kt when assuming typical intragalactic supernova-Earth distances. The few reactions with $\bar{\nu}_\mathrm{e}$ and ${\nu}_{\mu}$ in this material can be neglected in comparison to $\nu_\mathrm{e}$ related events. Moreover, natural lead has a comparatively small thermal neutron capture cross section of $\approx 0.15$ barn, which allows lead to moderate fast neutrons almost without absorption prior to arrival at a neutron counter \cite{Bol.12}. 
Although lead isotopes have large neutron excess, some of the neutrons might still be captured by $^{206}$Pb and $^{207}$Pb, though \cite{Kin.51}.

In comparison, a $^{56}$Fe target is characterized by somewhat weaker response to CC reactions with neutrinos (see Tab.~\ref{table2} and Tab.~\ref{table3} for the accretion and cooling phase neutrinos, respectively). The predicted total cross section of neutrino-induced reactions in $^{56}$Fe is 
$\approx 10$ times smaller than for $^{208}$Pb.
Almost half of the predicted events come from single-neutron emission from $^{56}$Co, while two-neutron emissions are 
severely reduced. 
Due to small neutron excess ($N-Z=4$), the detector based
on $^{56}$Fe is also sensitive to antineutrino CC reactions with cross 
sections of $\approx 10^{-42}$cm$^2$, an order of magnitude smaller than for $\nu_\mathrm{e}-^{56}$Fe reactions, but the daughter nuclei mostly deexcite by emission of $\gamma$ rays. 
Furthermore, the cross sections for absorption of thermal neutrons in natural iron is an order of magnitude larger than in lead ($\approx 2.5$ barn \cite{Bol.12}), and thus iron appears to be an inappropriate target material for a $\nu_\mathrm{e}$ and $\bar{\nu}_\mathrm{e}$ detector.

For mineral oil (water) with a density of $0.85$ g/cm$^3$ (1.0 g/cm$^3$ at 4 $^\circ$C), in which
the target nuclei are $^{12}$C ($^{16}$O), the predicted cross sections are of the order of $\approx 10^{-42}$ cm$^{2}$. Due to $N=Z=6$ $(N=Z=8)$, the difference in total number of primary events between neutrino and antineutrino CC reactions is smaller than for $N>Z$ nuclei.
Due to relatively high energy thresholds ($\gtrsim11 \text{ MeV}$) for both types of reactions in $^{12}$C ($^{16}$O), the expected responses are rather low for 
$\nu_{e}$ ($\bar{\nu}_{e}$) energies of $\lesssim20 \text{ MeV}$. 
Actually, in these detectors $\bar{\nu}_\mathrm{e}$ induce reactions mainly with 
free protons (see Tab.~\ref{table2} and Tab.~\ref{table3}), with a low energy threshold (1.8 MeV) and cross sections $\sigma\approx 10^{-43}(E_{\nu}/\mathrm{MeV})^{2}$.
The dominance of $\bar{\nu}_\mathrm{e}$-p
reactions in mineral oil and water ensures an efficient coverage 
of the $\bar{\nu}_\mathrm{e}$ spectra. 
Therefore, detectors based either on mineral oil or water can cover the antineutrino response in the hybrid method. In these cases the influence of CC reactions of $\nu_\mathrm{e}$ and $\bar{\nu}_\mathrm{e}$ with nuclei amounts only $\approx 1\%$ in the response and thus CC events with nuclei can be neglected. Similar results were obtained in Ref.~\cite{Bur.92}, i.e., only $\approx 1.5\%$($5.6\%$) of the reactions are expected to be related to $^{16}$O($^{12}$C) in CC and NC reactions in the KII detector (Large Volume Detector) in the energy range of interest. An additional $\gamma$ signal related to neutron capture in the detector (either on the original target nuclei or after adding an impurity such as Gd to increase the capture rates) can be used 
to confirm or eliminate charge exchange $\bar{\nu}_\mathrm{e}$-p events ~\cite{Apo.03}.

\subsection{\label{sec:application}Application of the hybrid method}
In the following we present calculations of the ratios of $\bar{\nu}_e$ and $\nu_e$ detector events for a variety of possible incoming supernova (anti)neutrino accretion and cooling phase fluxes (see Tab.~\ref{table1}), both for normal (NH) and inverted (IH) mass hierarchy.
By using $r_{\kappa_\mathrm{A} \kappa_\mathrm{B}}^\mathrm{AB}$ as defined in Eq.~(\ref{reducedc}), we obtained mass and distance invariant values for the most sensitive decay channels in the detectors. Statistical uncertainties in $r_{\kappa_\mathrm{A} \kappa_\mathrm{B}}^\mathrm{AB}$, however, still depend on such quantities as distance, etc.
Due to the reasons given above, we eliminated $^{56}$Fe from the further analysis
and used only $^{208}$Pb and mineral oil (water) for the neutrino and antineutrino sectors, respectively.
Including only dominant channels in the analysis, the quantity $r_{\kappa_\mathrm{A} \kappa_\mathrm{B}}^\mathrm{AB}$
should be the same in the Pb/H$_2$O and Pb/CH$_2$ detector combinations, due to the similarity of structural (empirical) formula of water and 
mineral oil with 2 free protons each.

In Fig.~\ref{hybrid_acc} the following ratios $r_{\kappa_\mathrm{A} \kappa_\mathrm{B}}^\mathrm{AB}$ of the detector 
events are shown for the incoming (anti)neutrino fluxes
of the accretion phase for NH and IH, including statistical uncertainties:
(a) $r_\mathrm{e^{+}1n}^\mathrm{free~p,Pb} $ (b) $r_\mathrm{e^{+}2n}^\mathrm{free~p,Pb}$, and (c) $r_\mathrm{e^{+}tot~n}^\mathrm{free~p,Pb}$.
The quantities (a), (b), and (c) correspond to the ratios between the numbers of positron events in the 
antineutrino detector (with dominant $\bar{\nu}_\mathrm{e}$-p events) and  neutrino-induced one-, two-, and total 
neutron emissions in the Pb detector, respectively.
%
We employed three combinations of the 
luminosities for the accretion phase (A1-A3) for three different configurations of initial average energies of neutrino spectra, as specified in Tab.~\ref{table1}.

As seen in Fig.~\ref{hybrid_acc}, the ratios of $\bar{\nu}_\mathrm{e}$ and $\nu_\mathrm{e}$ induced events result in
reasonable separation between NH and IH for the A2 case, which has a larger asymmetry in relative luminosities between electron and non-electron species, independent of initial average neutrino energies. The other cases of neutrino fluxes are characterized by partial overlaps between NH and IH event ratios for all energy configurations.
By considering other ratios shown in Fig.~\ref{hybrid_acc}, we note that neutrinos 
from the accretion phase systematically do not provide conclusive information to determine the
mass hierarchy, with the exception of flux $A2$.

In Fig.~\ref{hybrid_cool} we show the same decay channel analysis as in Fig.~\ref{hybrid_acc}, 
but for the neutrino fluxes of the supernova cooling phase.
Three types of the cooling phase (anti)neutrino fluxes were considered (C1-C3), as given
in Tab.~\ref{table1}, and the same energy configurations as for the accretion fluxes given above.
As seen in the figure, the (anti)neutrino induced events for the cooling phase are characterized
by complete separation of the ranges of values calculated for NH and IH: (a) $r_\mathrm{e^{+}1n}^\mathrm{free~p,Pb}$, 
(b) $r_\mathrm{e^{+}2n}^\mathrm{free~p,Pb}$, and (c) $r_\mathrm{e^{+}tot~n}^\mathrm{free~p,Pb}$ for all cases of supernova neutrino fluxes. 
Since the detectors with target material based on heavy nuclei are usually used only as counters, the number of emitted neutrons can provide additional information for the $\nu_\mathrm{e}$ part of spectra. Nevertheless, the strong dependence of e$^+$, one- and two-neutron emissions on the average initial energies of $\nu$ can be used to constrain their initial energy configurations (see Tab.\ \ref{table3}).

The results of the present analysis also display sensitivity to the degree of asymmetry of the initial relative luminosities between $\nu$ species. In the case of C2 (larger $l_\nu$ asymmetry), the $r$-quantities are systematically shifted toward smaller values (see Fig.\ \ref{hybrid_cool}). In the case of two-neutron emissions, a complete distinction of the $r_\mathrm{e^+, 2n}^\mathrm{free~p,Pb}$ quantities between two hierarchies is obtained, similar as in the case of one-neutron emissions. Therefore, in a detector with sensitivity to one- and two-neutron emissions, such as HALO, both types of the events can be used to constrain the neutrino mass hierarchy. 
We note that while this may be feasible for the (ii) and (iii) cases of supernova neutrino energy configurations,
in the (i) case the number of two-neutron events is rather small (see Tab.\ III). Thus, it would be difficult to provide reliable limits. Nevertheless, the separation of the $r$-quantities as shown in Fig.\ \ref{hybrid_cool}, especially in the case of a reasonable number of single-neutron and total neutron emission events (Tab.\ III), clearly demonstrates the promise in using the hybrid method to constrain the neutrino mass hierarchy. 
Additional, currently not available, information on the sensitivity and efficiency of the HALO detector would allow a further, more detailed  feasibility study for the proposed method.

\section{\label{sec:conclusion}Conclusion}
We have presented a hybrid method to determine the neutrino mass
hierarchy by simultaneous measurements of supernova $\nu_\mathrm{e}$ and 
$\bar{\nu}_\mathrm{e}$ events
in detectors based on different types of target material.
Using supernova $\nu_\mathrm{e}$($\bar{\nu}_\mathrm{e}$) fluxes that include
 collective and MSW effects for the 
accretion and cooling phases, responses in mineral oil, water, $^{56}$Fe and
$^{208}$Pb have been analyzed both for normal and inverted neutrino mass hierarchies.
The analysis of charge-exchange excitation spectra for
$^{56}$Fe and $^{208}$Pb targets demonstrates the sensitivity of the nuclear response 
to the neutrino mass hierarchy, that is different for various multipole transitions.
The hybrid method, that combines antineutrino-induced events
in water or mineral oil with neutrino-induced emissions in heavier target nuclei 
(such as $^{208}$Pb), could provide a useful tool to constrain the neutrino
mass hierarchy. The number of 
emitted neutrons for NH in general is larger than for IH, both
for $^{208}$Pb and $^{56}$Fe target nuclei. Since it is rather difficult to compare
the absolute values of calculated particle emissions with the detector events, 
a set of hierarchy dependent relative quantities is introduced, which are independent 
of the supernova-Earth distance, of the effective number of targets in $\nu_\mathrm{e}$ 
and $\bar{\nu}_\mathrm{e}$ detectors, of their total masses, and of the ratios of these quantities. 
They contain the information on nuclear structure, excited states, relevant charge-exchange
transitions and particle emission modes induced in nuclei by supernova (anti)neutrinos.
The incoming (anti)neutrino fluxes employed in this study include the imprints of collective and MSW effects during the ccSN event and these are generally different between NH and IH. Model calculations show that the ratio of the numbers of $\nu_\mathrm{e}$ and $\bar{\nu}_\mathrm{e}$ induced detector events represents a quantity that allows to distinguish between the two neutrino mass hierarchies. Simultaneous comparison of several ratios of the numbers of 
detector events associated to various emitted 
particles for different supernova phases would reduce the uncertainties that may arise by considering only a single quantity.

As confirmed in calculations, heavy nuclei are almost completely inert to CC reactions 
with antineutrinos due to the strong Pauli blocking of neutron single-particle states. The pure neutrino signal and 
large CC cross section for reactions with neutrinos establish $^{208}$Pb as probably the most 
important nuclear target for neutrino detection and reconstruction of the
neutrino part of spectra. On the other side, detectors based on mineral oil or water as target-material represent the most reasonable choice for $\bar{\nu}_e$ detection.
Taking into account current neutrino detector developments (in particular, HALO ($^{208}$Pb)~\cite{Dub.08}, Super-Kamiokande (water)~\cite{Kam.14}, etc.), the hybrid method presented in this work will provide a useful tool to constrain the neutrino mass hierarchy when the next galactic ccSN appears.

\begin{acknowledgments}
We thank Kate Scholberg, Meng-Ru Wu, and Maik Frensel for the information and comments
of relevance for the paper. This work has been supported in part by
FP7-PEOPLE-2011-COFUND program NEWFELPRO, the Swiss National 
Science Foundation, the Croatian Science Foundation under the
project Structure and Dynamics of Exotic Femtosystems (IP-2014-09-9159), 
and by Deutscher Akademischer Austausch Dienst (DAAD). T.R. acknowledges support from the European Research Council (grant GA 321263-FISH) and the British STFC grant ST/M000958/1. 
\end{acknowledgments}

%
\newpage
\begin{table}
\caption{{Initial lumininosity ratios of (anti)neutrino species for type II supernova spectra in accretion(A) and cooling(C) phase. Three possible combinations of the luminosity ratios for (anti)neutrino species are shown for each phase, together with fixed total energy emitted by (anti)neutrinos and time duration of the phases.}}
\begin{center}
\item[]\begin{tabular}{@{}*{3}{l}}
\hline\hline\cr
$ $&$l_{{\nu}_\mathrm{e}}:l_{\bar{\nu}_\mathrm{e}}:l_{{\nu}_x} $&$T_{int}$\cr
\hline\cr
Accretion ph.&$\approx 0.4 \times 10^{53}$ erg& \cr\cr
A1&$4.0~:~4.0~:~3.0$&0.4 s\cr
A2&$3.0~:~3.0~:~1.0$&\cr
A3&$12.0~:~8.0~:~5.0$&\cr
\hline
Cooling ph.&$\approx 2\times10^{53}$ erg& \cr\cr
C1&$1.0~:~1.0~:~1.0$&10.0 s\cr
C2&$2.0~:~2.0~:~4.0$&\cr
C3&$3.2~:~2.8~:~3.5$&\cr
\hline\cr
Total&$\approx 2.4\times 10^{53}$ erg& \cr
\hline\hline

\end{tabular}
\end{center}
\label{table1}
\end{table}

\newpage

\begin{table}
\caption{{Detector response for $\nu_\mathrm{e}$($\bar{\nu}_\mathrm{e}$)-induced reactions in mineral oil (CH$_{2}$), water (H$_{2}$O), $^{56}$Fe and $^{208}$Pb,
 for the incoming (anti)neutrino fluxes of the accretion supernova phase, both for
normal (NH) and inverted (IH) neutrino mass hierarchies.
Only dominant emission channels are shown, including e$^+$, one- and two-neutron emissions.
A1, A2, A3 denote the combinations of the luminosity ratios of (anti)neutrino
species given in Tab.~\ref{table1}, while (i) stands for $\langle E_{\nu_\mathrm{e}}^0 \rangle$ = 8 MeV, $\langle E_{\bar{\nu}_\mathrm{e}}^0 \rangle$ = 11 MeV, $\langle E_{\nu_x}^0 \rangle$ = 13 MeV; (ii) is $\langle E_{\nu_\mathrm{e}}^0 \rangle$ = 10 MeV, $\langle E_{\bar{\nu}_\mathrm{e}}^0 \rangle$ = 13 MeV, $\langle E_{\nu_x}^0 \rangle$ = 15 MeV; and (iii) is $\langle E_{\nu_\mathrm{e}}^0 \rangle$ = 12 MeV, $\langle E_{\bar{\nu}_\mathrm{e}}^0 \rangle$ = 15 MeV, $\langle E_{\nu_x}^0 \rangle$ = 19 MeV. }}
\begin{center}
\item[]\begin{tabular}{@{}*{8}{l}}
\hline\hline\cr
Accretion ph.&L type&\multicolumn{2}{c}{(i)} & \multicolumn{2}{c}{(ii)} &\multicolumn{2}{c}{(iii)} \cr
0.4 s&$ $&\multicolumn{2}{c}{NH$~~~~~~~~~$IH}&\multicolumn{2}{c}{NH$~~~~~~~~~$IH}&\multicolumn{2}{c}{NH$~~~~~~~~~$IH}\cr
\hline\hline\cr
p($\bar{\nu}_\mathrm{e}$,e$^+$)n in CH$_{2}$\cr
e$^+$&A1&45~-~60&42~-~56&54~-~70&49~-~64&65~-~82&61~-~78\cr
$ $&A2&58~-~74&28~-~39&69~-~87&32~-~45&82~-~101&41~-~55\cr
$ $&A3&43~-~57&34~-~47&51~-~67&40~-~54&61~-~78&51~-~66\cr
\hline\cr
p($\bar{\nu}_\mathrm{e}$,{e}$^+$)n in H$_{2}$O\cr
e$^+$&A1&34~-~47&32~-~44&41~-~55&37~-~50&49~-~65&47~-~62\cr
$ $&A2&44~-~58&21~-~31&53~-~69&25~-~36&62~-~79&31~-~43\cr
$ $&A3&33~-~45&26~-~37&39~-~53&30~-~43&47~-~61&39~-~52\cr
\hline\cr
$^{56}$Fe($\nu_\mathrm{e}$,e$^-$)$^{56}$Co\cr
1n&A1&0~-~4&0~-~3&2~-~6&2~-~6&6~-~13&6~-~13\cr
$ $&A2&0~-~2&0~-~2&1~-~4&1~-~4&4~-~9&4~-~9\cr
$ $&A3&0~-~3&0~-~3&1~-~5&1~-~5&5~-~11&5~-~11\cr
\hline\cr
1n&A1&12~-~20&12~-~20&19~-~29&19~-~29&35~-~48&35~-~48\cr
$ $&A2&7~-~14&7~-~14&12~-~20&12~-~20&22~-~33&22~-~33\cr
$ $&A3&10~-~17&10~-~17&16~-~25&16~-~25&29~-~41&29~-~41\cr\cr
2n&A1&0~-~3&0~-~3&1~-~5&1~-~5&7~-~14&7~-~14\cr
$ $&A2&0~-~2&0~-~2&0~-~4&0~-~4&4~-~10&4~-~10\cr
$ $&A3&0~-~2&0~-~2&1~-~5&1~-~5&6~-~12&6~-~12\cr\cr
total~n&A1&15~-~24&15~-~24&26~-~38&26~-~38&56~-~72&56~-~72\cr
$ $&A2&9~-~17&9~-~17&17~-~26&17~-~26&36~-~49&36~-~49\cr
$ $&A3&12~-~20&12~-~20&22~-~32&22~-~32&46~-~61&46~-~61\cr
\hline\hline\cr
\end{tabular}
\end{center}
\label{table2}
\end{table}

\newpage

\begin{table}
\caption{The same as Tab.~\ref{table2} but for the cooling supernova phase.
C1, C2, C3 denote the combinations of the luminosity ratios for (anti)neutrino species
given in Tab.~\ref{table1}.}
\begin{center}
\item[]\begin{tabular}{@{}*{8}{l}}
\hline\hline\cr
Cooling ph.&L type&\multicolumn{2}{c}{(i)} & \multicolumn{2}{c}{(ii)} &\multicolumn{2}{c}{(iii)} \cr
10 s&$ $&\multicolumn{2}{c}{NH$~~~~~~~~~$IH}&\multicolumn{2}{c}{NH$~~~~~~~~~$IH}&\multicolumn{2}{c}{NH$~~~~~~~~~$IH}\cr
\hline\hline\cr
p($\bar{\nu}_\mathrm{e}$,e$^+$)n in CH$_{2}$\cr
e$^+$&C1&253~-~285&248~-~281&303~-~339&295~-~331&394~-~435&370~-~409\cr
$ $&C2&217~-~248&211~-~242&270~-~304&257~-~291&378~-~418&340~-~378\cr
$ $&C3&235~-~267&230~-~262&286~-~320&276~-~310&381~-~421&352~-~391\cr
\hline\cr
p($\bar{\nu}_\mathrm{e}$,{e}$^+$)n in H$_{2}$O\cr
e$^+$&C1&195~-~224&192~-~220&234~-~266&228~-~259&305~-~341&286~-~321\cr
$ $&C2&167~-~194&163~-~190&209~-~238&199~-~228&292~-~327&263~-~296\cr
$ $&C3&181~-~209&178~-~205&220~-~251&213~-~243&294~-~330&272~-~306\cr
\hline\cr
$^{56}$Fe($\nu_\mathrm{e}$,e$^-$)$^{56}$Co\cr
1n&C1&11~-~19&7~-~14&21~-~31&13~-~21&46~-~60&27~-~39\cr
$ $&C2&14~-~23&9~-~16&25~-~36&15~-~24&55~-~71&30~-~42\cr
$ $&C3&12~-~20&7~-~14&22~-~32&14~-~22&48~-~63&28~-~40\cr\cr
\hline\cr
$^{208}$Pb($\nu_\mathrm{e}$,e$^-$)$^{208}$Bi& & & & & &\cr
1n&C1&88~-~108&68~-~85&126~-~150&95~-~116&194~-~223&139~-~163\cr
$ $&C2&106~-~128&82~-~101&151~-~177&111~-~133&230~-~262&154~-~180\cr
$ $&C3&93~-~113&71~-~89&133~-~157&99~-~121&204~-~234&144~-~169\cr\cr
2n&C1&9~-~16&4~-~9&19~-~29&10~-~17&53~-~69&28~-~40\cr
$ $&C2&11~-~18&5~-~11&23~-~34&11~-~19&63~-~80&31~-~44\cr
$ $&C3&9~-~16&4~-~10&20~-~30&10~-~18&56~-~72&29~-~41\cr\cr
total~n&C1&112~-~134&82~-~101&173~-~201&122~-~145&313~-~350&205~-~235\cr
$ $&C2&135~-~159&98~-~119&207~-~237&141~-~166&371~-~410&228~-~259\cr
$ $&C3&118~-~141&86~-~105&182~-~210&128~-~151&329~-~366&213~-~243\cr
\hline\hline\cr
\end{tabular}
\end{center}
\label{table3}
\end{table}

\newpage

\begin{figure}
\includegraphics[scale=0.60,angle=0]{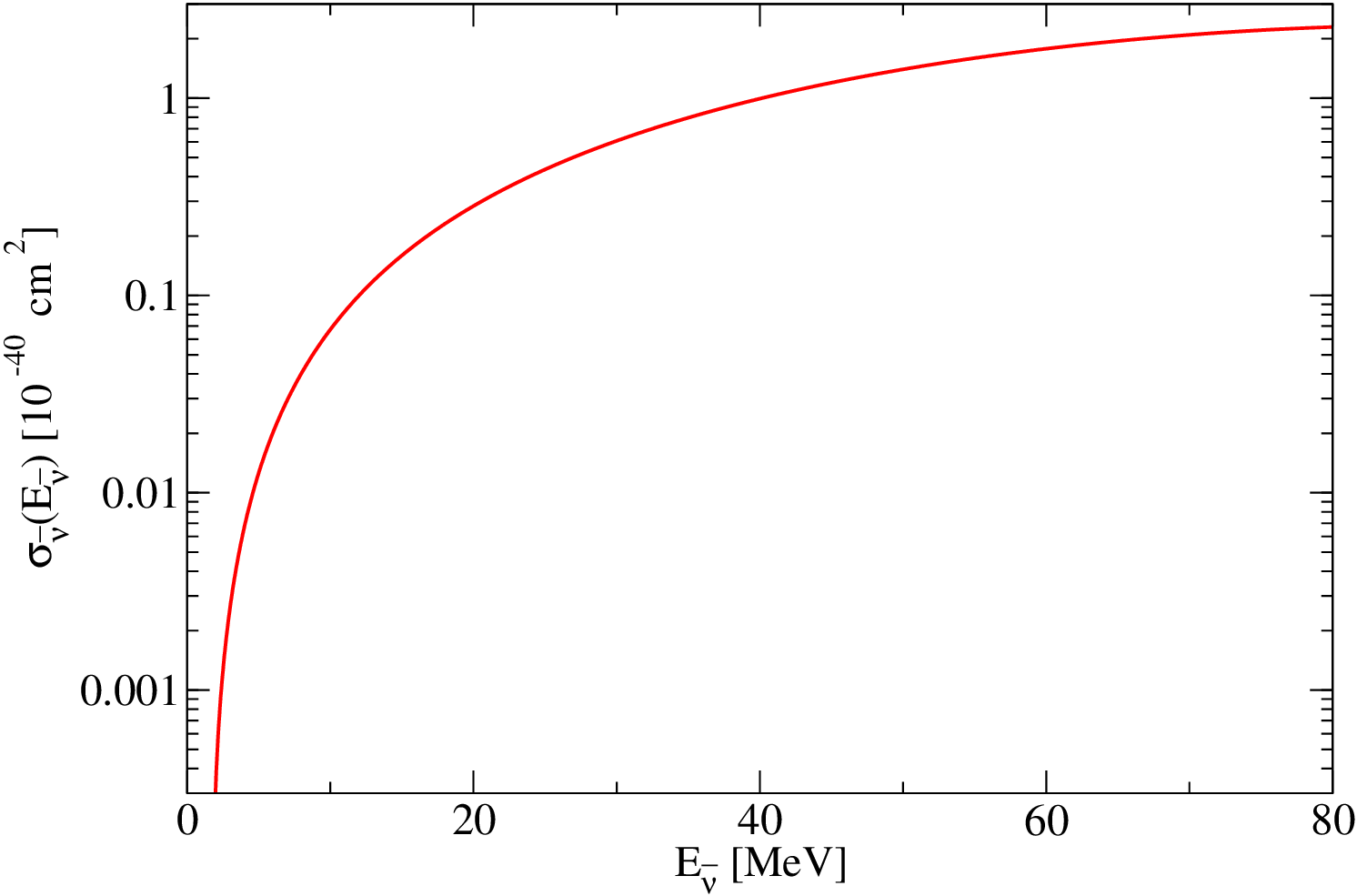}
\caption{The cross sections for the inelastic scattering of $\bar{\nu}_{e}$ on free protons, shown as a function of the incoming antineutrino energy. Calculations are based on heavy-baryon
chiral perturbation theory  with radiative corrections from Ref.~\cite{Rah.12, Rah.12e}.}
\label{anu_proton}
\end{figure}

\newpage

\begin{figure}
\includegraphics[scale=0.6,angle=0]{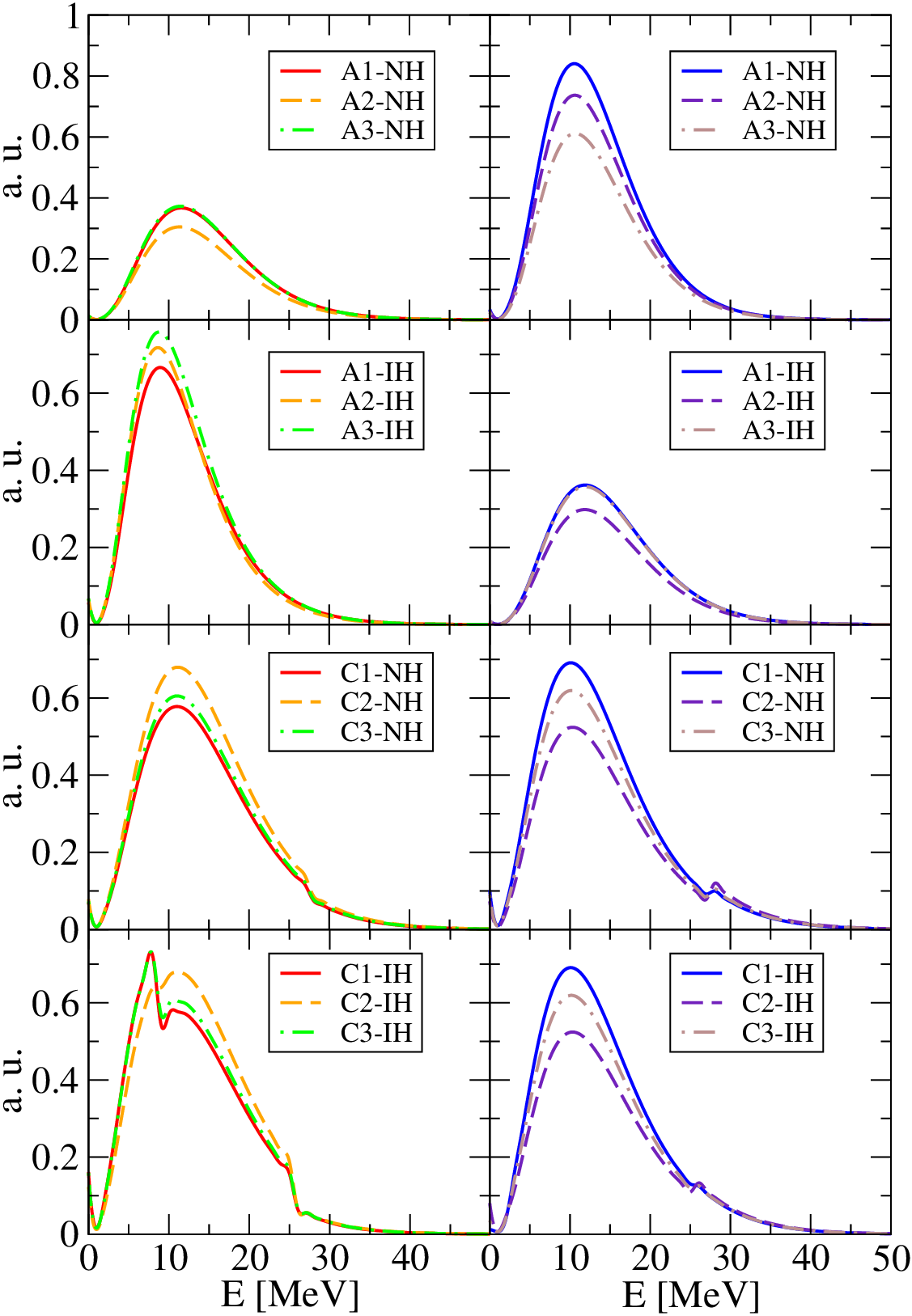}
\caption{Incoming $\nu_\mathrm{e}$ (left side) and $\bar{\nu}_\mathrm{e}$ fluxes (right side) 
for the accretion (A) and cooling (C) phase of core-collapse supernovae as a function
of neutrino energy for normal (NH) and inverted neutrino mass hierarchy (IH), for initial $\nu$ energy configuration (ii).
Three combinations of the luminosity ratios for (anti)neutrino species used in the accretion (A1-3)
and cooling (C1-3) phases are given in Tab.~\ref{table1}.
}
\label{fluxes}
\end{figure}

\newpage

\begin{figure}
\includegraphics[scale=0.60,angle=0]{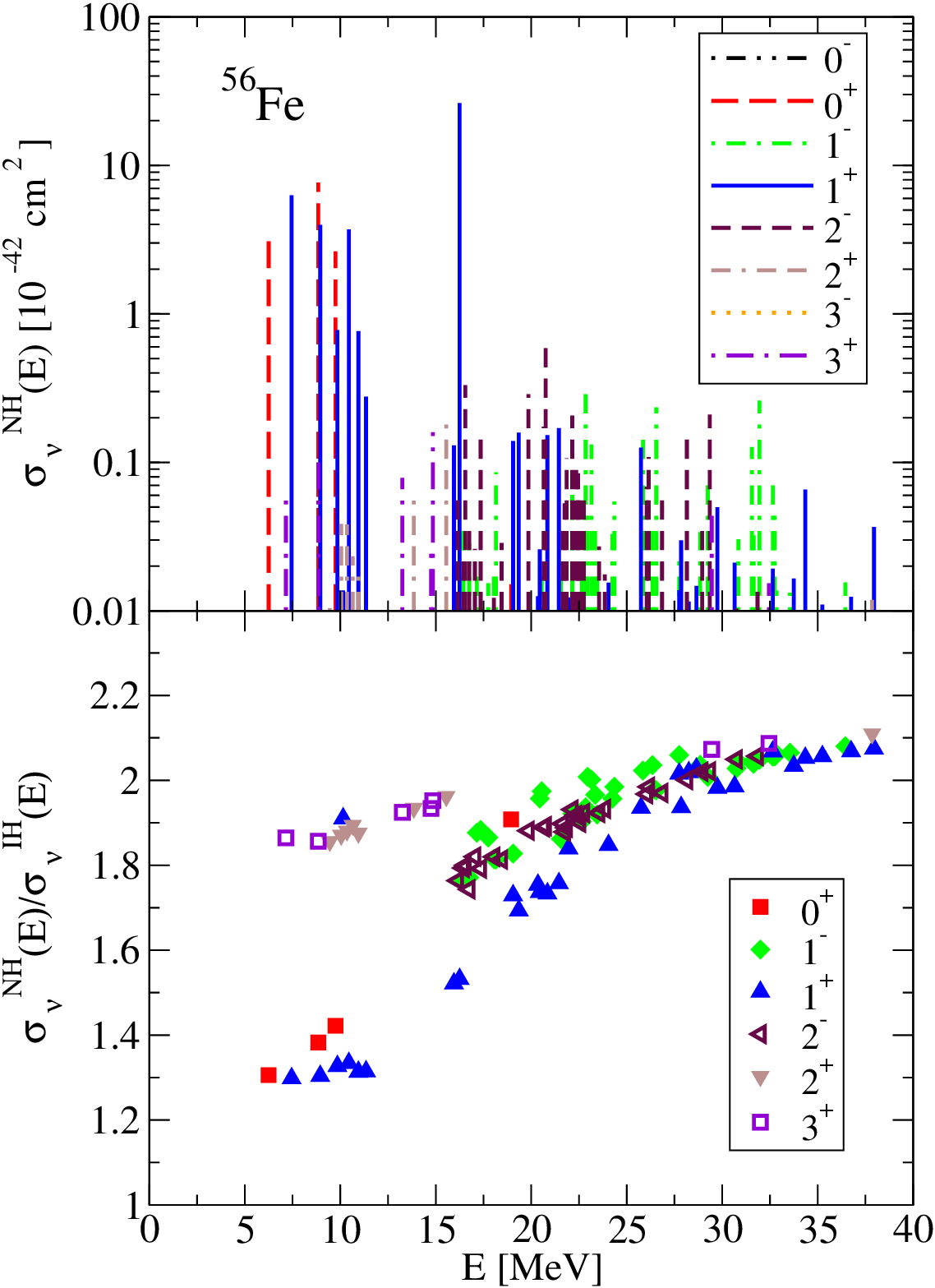}
\caption{Multipole decomposition of the flux averaged cross sections of the $\nu_\mathrm{e}$-$^{56}$Fe reaction as a function of excitation energy in the initial nucleus, calculated for the incoming cooling phase neutrino spectra (C3) in normal hierarchy (NH), including multipoles $J^{\pi}=0^{\pm}-3^{\pm}$ (upper panel). The ratio of cross sections between the normal (NH) and inverted (IH) neutrino mass hierarchies is shown in the lower panel.}
\label{kombo_fe}
\end{figure}

\newpage

\begin{figure}
\includegraphics[scale=0.60,angle=0]{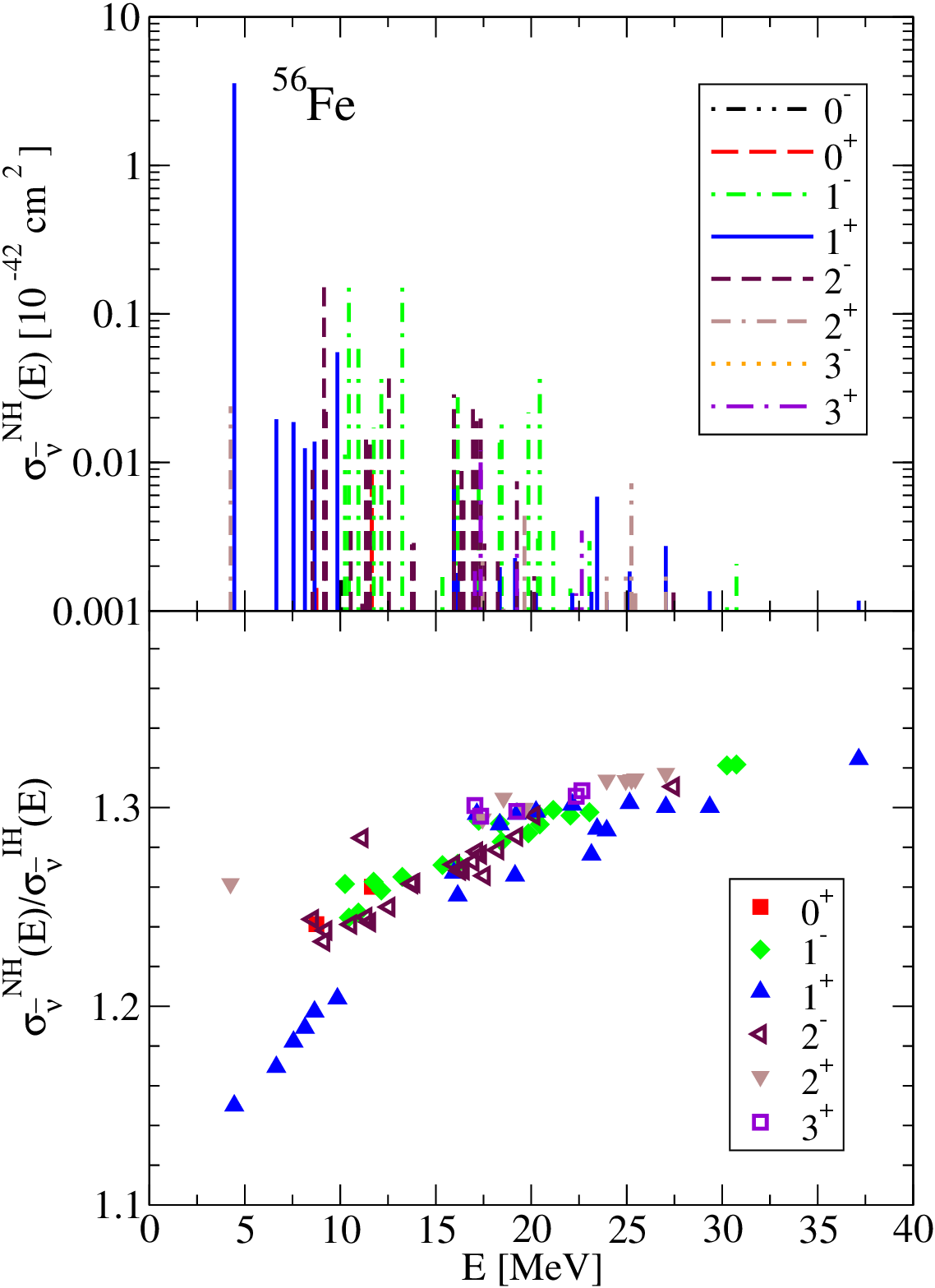}
\caption{Same as Fig.~\ref{kombo_fe}, but for the $\bar{\nu}_\mathrm{e}$-$^{56}$Fe reaction.}
\label{kombo_afe}
\end{figure}

\newpage

\begin{figure}
\includegraphics[scale=0.60,angle=0]{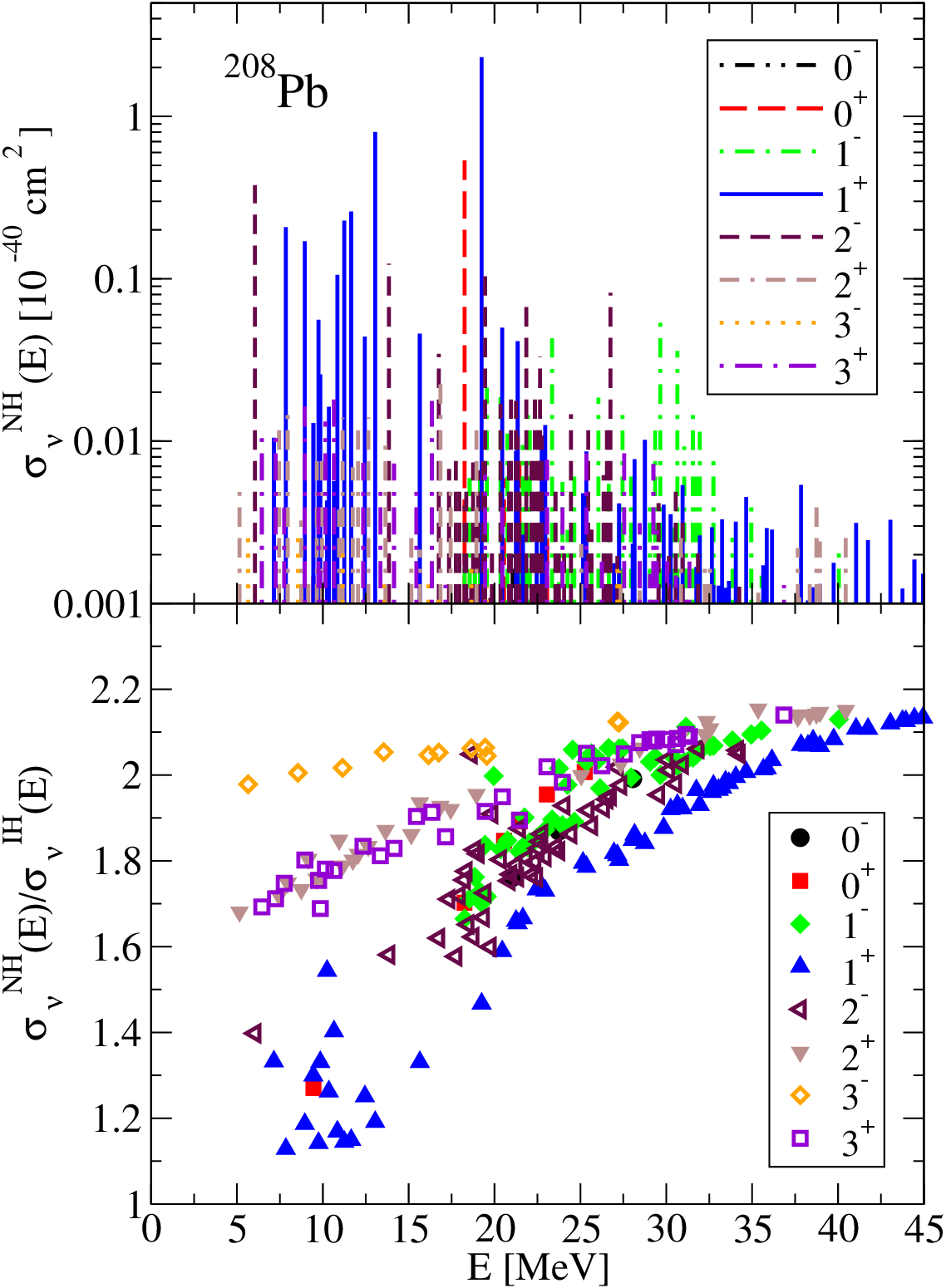}
\caption{Same as Fig.~\ref{kombo_fe}, but for the $\nu_\mathrm{e}$-$^{208}$Pb reaction.}
\label{kombo_pb}
\end{figure}

\newpage

\begin{figure}
\includegraphics[scale=0.6,angle=0]{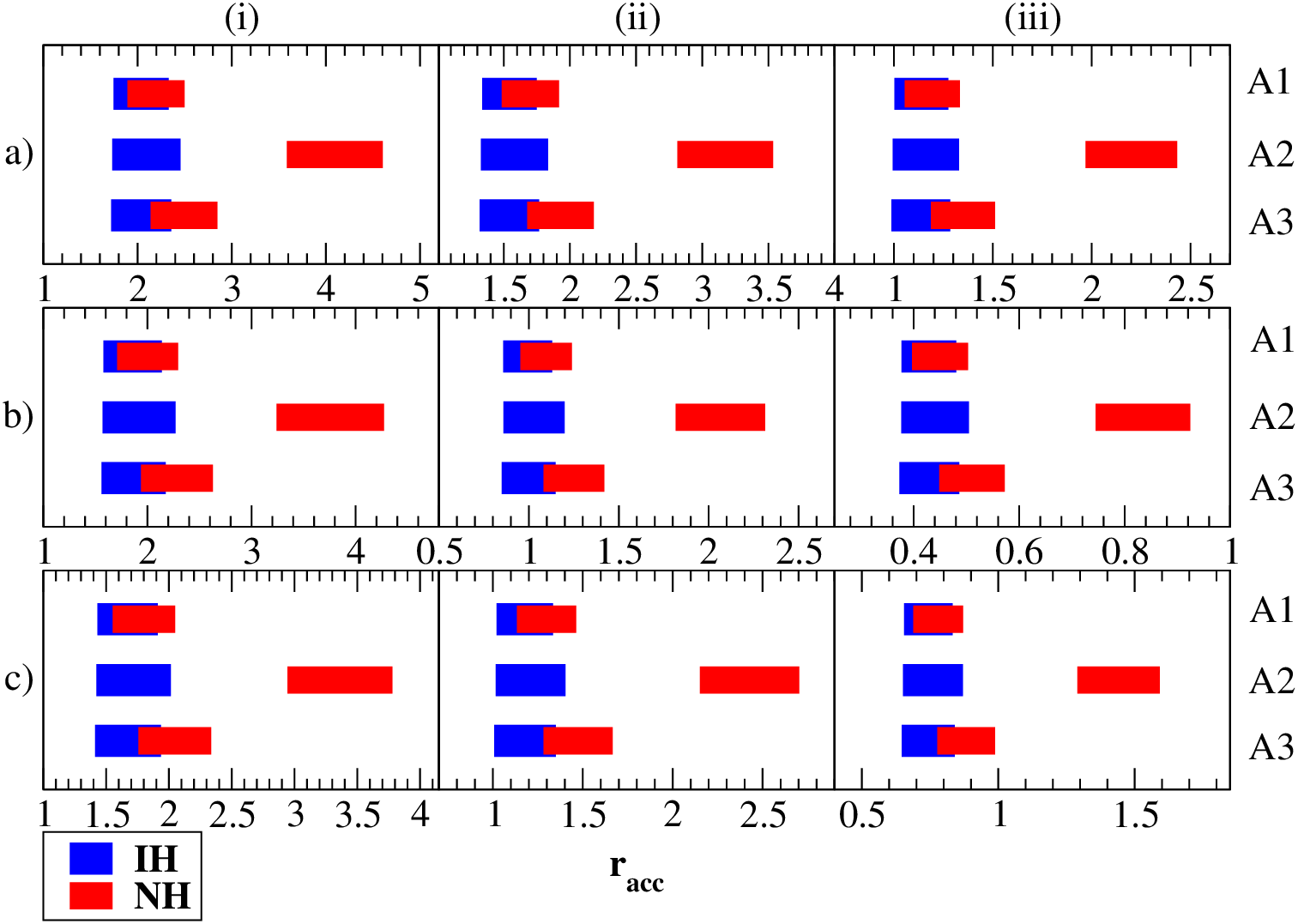}
\caption{
The ratios $r_{acc}$ of the number of the detector events induced in mineral oil (CH$_{2}$), water (H$_{2}$O), and $^{208}$Pb for the incoming (anti)neutrino fluxes
of the accretion phase in normal (NH) and inverted (IH) mass hierarchy: (a) $r_\mathrm{e^{+},1n}^\mathrm{free~p,Pb} \left[10^{-1}\right]$ (b) $r_\mathrm{e^{+},2n}^\mathrm{free~p,Pb}$ (c) $r_\mathrm{e^{+},tot~n}^\mathrm{free~p,Pb}\left[10^{-1}\right]$.
The three columns correspond to (i) $\langle E_{\nu_\mathrm{e}}^0 \rangle$ = 8 MeV, $\langle E_{\bar{\nu}_\mathrm{e}}^0 \rangle$ = 11 MeV, $\langle E_{\nu_x}^0 \rangle$ = 13 MeV; (ii) $\langle E_{\nu_\mathrm{e}}^0 \rangle$ = 10 MeV, $\langle E_{\bar{\nu}_\mathrm{e}}^0 \rangle$ = 13 MeV, $\langle E_{\nu_x}^0 \rangle$ = 15 MeV; and (iii) $\langle E_{\nu_\mathrm{e}}^0 \rangle$ = 12 MeV, $\langle E_{\bar{\nu}_\mathrm{e}}^0 \rangle$ = 15 MeV, $\langle E_{\nu_x}^0 \rangle$ = 19 MeV. A1, A2, A3 denote the combinations of the luminosity ratios for the
accretion phase given in Tab.~\ref{table1}.
}
\label{hybrid_acc}
\end{figure}

\newpage

\begin{figure}
\includegraphics[scale=0.6,angle=0]{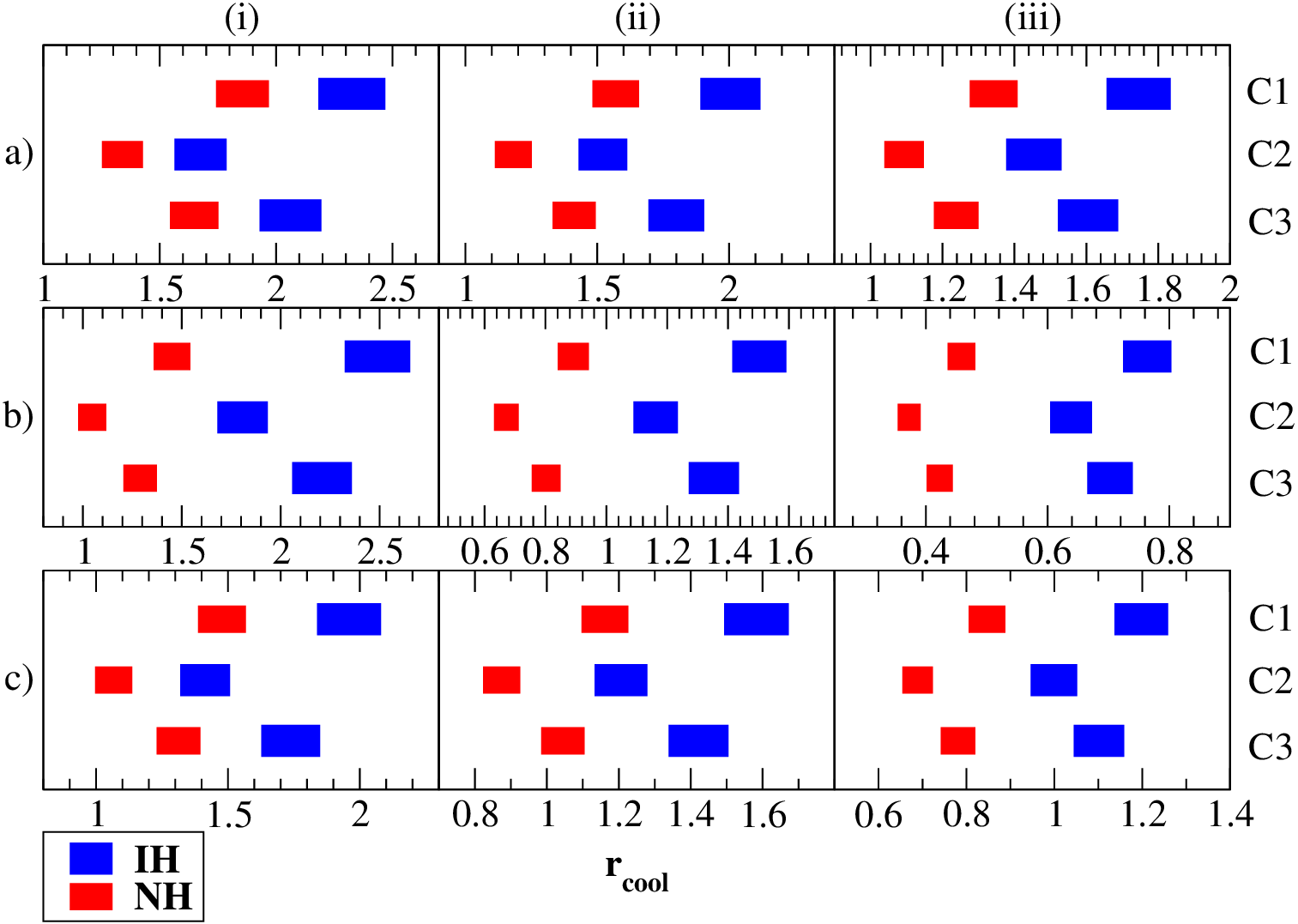}
\caption{
The ratios $r_{cool}$ of the number of the detector events induced in mineral oil (CH$_{2}$), water (H$_{2}$O), and $^{208}$Pb for the incoming (anti)neutrino fluxes
of the cooling phase in normal (NH) and inverted (IH) mass hierarchy for three configurations of initial $\nu$ average energies. The same notation as
in Fig.~\ref{hybrid_acc} applies. C1, C2, C3 denote the combinations of the luminosity ratios for the cooling phase
given in Tab.~\ref{table1}.}

\label{hybrid_cool}
\end{figure}

\end{document}